\documentclass[12pt]{iopart}

\usepackage{subcaption}
\usepackage{graphicx}
\expandafter\let\csname equation*\endcsname\relax
\usepackage{float}
\expandafter\let\csname endequation*\endcsname\relax
\usepackage{amsmath}
\usepackage{hyperref}
\usepackage[compress]{cite}

\usepackage{xcolor}
\usepackage{soul}

\usepackage[table]{xcolor}
\usepackage{geometry}
\usepackage{array}
\usepackage{booktabs}

\begin{document}

\title[]
{\textbf{Functional renormalization group study of $\rho$ meson
condensate at a finite isospin chemical potential in the quark meson model}}
\author{Mohammed Osman\textsuperscript{*\,$1$}, Defu Hou\textsuperscript{\dag\,$1$},Wentao Wang\textsuperscript{\ddag\,$1$} and Hui Zhang\textsuperscript{\S\,$2$,$3$,$4$}}

\address{$^1$Institute of Particle Physics (IOPP) and key laboratory of Quark and lepton physics
(MOE), Central China Normal University, Wuhan 430079, China}\

\address{$^2$State Key Laboratory of Nuclear Physics and Technology, Institute of Quantum Matter, South China Normal University, Guangzhou 510006, China}\

\address{$^3$Guangdong Basic Research Center of Excellence for Structure and Fundamental Interactions of Matter, Guangdong Provincial Key Laboratory of Nuclear Science, Guangzhou 510006, China}\ 

\address{$^4$Physics Department and Center for Exploration of Energy and Matter, Indiana University, 2401 N Milo B. Sampson Lane, Bloomington, IN 47408, USA}


\ead{mhdosmn183@mails.ccnu.edu.cn}
\ead{houdf@mail.ccnu.edu.cn}
\ead{wwt@mails.ccnu.edu.cn}
\ead{Mr.zhanghui@m.scnu.edu.cn}
\vspace{10pt}
\begin{indented}
\item[]
\end{indented}
\begin{abstract}
We investigated the effect of an isospin chemical potential ($\mu_{I}$) within the quark-meson model, which approximates quantum chromodynamics (QCD) by modeling low-energy phenomena such as chiral symmetry breaking and phase structure under varying conditions of temperature and chemical potential. Using the functional renormalization group (FRG) flow equations, we calculated the phase diagram in the chiral limit within the two-flavor quark-meson model in a finite $\mu_{I}$ with $\rho$ vector meson interactions.
Fluctuation effects significantly decrease the critical chemical potential from the mean-field (MF) value $\mu_{I, MF} > m_\rho$ to a lower value, at which point the $\rho$ vector meson condensates alongside the chiral condensate once the isospin chemical potential exceeds the critical value $\mu_{I}^{\text{crit}}$.
This $\rho$ condensation was investigated numerically for different meson coupling strengths. The $\rho$ meson dominated region is delineated from other phases by a second-order phase transition at lower $\mu_{I}$ and a first-order transition at slightly higher $\mu_{I}$.
\end{abstract}
%
%
%
%
%

\section{INTRODUCTION}

Quantum chromodynamics (QCD) describes the strong interactions between quarks and gluons. Its phase structure, dependent on temperature and chemical potential, reveals various phases of matter \cite{Meyer-Ortmanns:1996ioo,Rischke:2003mt,kogut2003phases,Suganuma2020,Fukushima:2010bq,Stephanov:2004wx}. The phase structure is often represented in the baryon chemical potential $\mu_{B}$ versus temperature $T$ plane, known as the $\mu_{B}$-$T$ plane \cite{Vuorinen:2003fs,McLerran:2007qj,Ipp:2006ij,Hands:2006ve,Braatenjr:1995}. This diagram illustrates the various phases of QCD matter under extreme conditions\cite{ADAMI19931,Wilczek:1999ym,kogut2003phases,Blaschke:2018mqw,Huang:2023ogw}, such as those in high-energy collisions or dense astrophysical objects such as neutron stars.

The isospin chemical potential ($\mu_{I}$) significantly influences the structure of the QCD phase diagram by shifting the boundaries of critical transitions such as chiral symmetry restoration and quark-gluon plasma (QGP) formation to higher values of baryon chemical potential ($\mu_{B}$)\cite{CamaraPereira:2020xla,Detmold:2012wc,stieleab2012qcd,Brandt:2016zdy}. This is due to the isospin asymmetry between up and down quarks\cite{Kogut:2004zg}.
When $\mu_{I}$ exceeds the pion mass, a pion condensation phase occurs, resulting in spontaneous symmetry breaking \cite{Carignano:2016lxe,Brauner:2016lkh,Son:2000by,Son:2000xc}. Furthermore, large values of $\mu_{I}$ alter the formation of QGP and modify the behavior of color superconducting phases, especially in high-density environments such as neutron stars\cite{RevModPhys.80.1455}.
In this connection, the lattice (QCD) is free from the sign problems, thereby permitting the use of the of standard Monte Carlo theory to calculate thermodynamic properties and to draw the phase diagram as a function of temperature (T) and isospin ($\mu_{I}$). This allows  confrontation with low-energy effective theories, such as chiral perturbation theory \cite{MILC:2009mpl,Pich:1995bw,Bernard:2007zu}, as well as with models such as the Nambu-Jona-Lasinio \cite{PhysRev.122.345,Nambu:1961fr,HATSUDA1994221,PhysRevC.53.410,BUBALLA2005205,PhysRevD.94.094001,Fukushima:2012xw,NJL2,NJL3,NJL4} and quark-meson models\cite{Andersen:2013swa,Fu:2015naa,Jung:2016yxl,Herbst:2013ufa,Herbst:2013ail,Strodthoff:2011tz,Fukushima:2012xw,Kamikado:2012cp,Tripolt:2013jra}.

In Ref.~\cite{Aharony:2007uu}, holographic QCD was employed to investigate $\rho$-meson condensation, providing valuable insights from the strong-coupling regime that extend our understanding of this phenomenon beyond conventional approaches. In the present study, we addressed a related but simpler problem: $\rho$ condensation in the presence of a finite isospin chemical potential $\mu_{I}$. Using a two-flavor quark–meson model, we explored the formation and behavior of the $\rho$ condensate under varying values of $\mu_{I}$, temperature ($T$), and coupling constant ($g_{\rho}$).
 Our objective was to advance the understanding of QCD phase structure and the properties of dense hadronic matter. Calculations within such models are typically performed at the mean-field level; going beyond this approximation is an essential step toward a more complete and realistic description. To achieve this, we employed the functional renormalization group (FRG) method~\cite{WETTERICH199390,doi:10.1142/S0217751X94000972,Ellwanger:1993mw,Berges:2000ew,Gies:2006wv,Pawlowski:2005xe,Delamotte:2007pf}, which systematically incorporates quantum and thermal fluctuations beyond mean-field theory.
FRG is a potent non-perturbative method that allows quantum and thermal fluctuations to be incorporated into a field theory.
FRG has been extensively used to investigate the QCD phase diagram with chiral effective models beyond MF, such as the NJL and the QM models.

This paper is organized as follows. The 2-flavour quark-meson model, including the $\rho$ vector meson, and the FRG method are presented in Sect. II.
 In Sect. III, 
the results are discussed. Finally, in Sect. IV, 
conclusions are presented.

\section{QUARK-MESON MODEL WITH \texorpdfstring{$\rho$}{rho} VECTOR MESONS}\label{sec:2}

The Lagrangian of the two-flavour Quark-Meson model with $\rho$ vector meson in the Euclidean space is
\begin{eqnarray}
\mathcal{L} & =\bar{\psi}\left(\gamma_\mu \partial^\mu+\frac{\mu_I}{2} \gamma_0 \tau_3-\mu \gamma_0\right) \psi \nonumber \\
& -\bar{\psi}\left[g_s\left(\sigma+i \gamma_5 \boldsymbol{\tau} \cdot \boldsymbol{\pi}\right)+i \gamma_\mu\left(g_\rho \boldsymbol{\tau} \cdot \boldsymbol{\rho}^\mu\right)\right] \psi \nonumber \\
& +\frac{1}{2} \partial_\mu \sigma \partial^\mu \sigma+\frac{1}{2} \partial_\mu \boldsymbol{\pi} \partial^\mu \boldsymbol{\pi}+\frac{1}{4} \boldsymbol{R}_{\mu \nu}^{(\rho)} \boldsymbol{R}^{(\rho) \mu \nu} \nonumber \\
& -U(\sigma, \boldsymbol{\pi}, \boldsymbol{\rho}_{\mu}) .
\label{ Lagrangian}
\end{eqnarray}
 The field strength tensors of the vector bosons $\boldsymbol{\rho_\mu}$ are generally expressed as  $\boldsymbol{R}^{(\rho)}_{\mu\nu}=\partial_\mu \boldsymbol{\rho}_\nu-\partial_\nu \boldsymbol{\rho}_\mu-g_\rho \boldsymbol{\rho}_\mu \times \boldsymbol{\rho}_\nu$. A field $\psi$ is the light two-flavor quark field $=(u,d)^T$, and coupled to scalar $(\sigma)$ and isoscalar $(\boldsymbol{\pi})$ fields transforming as a four-component field $(\sigma,\boldsymbol{\pi})^T$ under the chiral group. A bold symbol stands for a vector. Here, $\boldsymbol{\tau}=(\tau_1,\tau_2,\tau_3)$ are the Pauli matrices in the isospin space, thereby introducing the isospin chemical potential $\mu_I=\mu_u-\mu_d$. Treated as background fields, only $\rho^3_0$ is non-vanishing, and the non-abelian of $\boldsymbol{R}^{(\rho)}_{\mu\nu}$ does not contribute in practice. The spatial components vanish because of the assumption of homogeneous isotropic matter. A Hubbard-Stratonovich transformation bosonizes these interactions, introducing effective vector-isovector fields, $\boldsymbol{\rho}_\mu$. While the fluctuations of the $\boldsymbol{\pi}$ and $\sigma$ fields are included non-perturbatively, $\boldsymbol{\rho}_\mu$ are treated as mean fields. These vector bosons conveniently parametrize unresolved short-distance physics. The potential for $\sigma,\boldsymbol{\pi}$ and $\boldsymbol{\rho}_{\mu}$ is
  \begin{equation}
\begin{array}{l}
\displaystyle U(\sigma, \boldsymbol{\pi}, \boldsymbol{\rho}_{\mu})=\frac{\lambda}{4}(\sigma^2+\boldsymbol{\pi}^2-f_{\pi}^2)^2-\frac{m^2_\rho}{2}\boldsymbol{\rho}_\mu \boldsymbol{\rho}^\mu,\\[5pt]
\end{array}
\end{equation}
where $f_\pi$  is the pion decay constant. We set $f_\pi=93$ MeV, and $m_\rho\sim 1$ GeV.
The parameters in our model are $g_s$, $g_\rho$ and $\lambda$. The values of these parameters for the FRG calculations are set to reproduce the same value for quantities such as a constituent quark mass of $\sim 300 MeV$. Regarding the value of $g_\rho$ and $m_\rho$, in our calculations, they always appear in the form of $g_\rho / m_\rho$. Therefore, their values are not discussed independently.
\subsection{FRG flow equation}
FRG is a powerful non-perturbative method that allows incorporating quantum and thermal fluctuations in a field theory~\cite{Berges:2000ew,Dupuis:2020fhh} and has been extensively applied to effective QCD models~\cite{Tripolt:2013jra,Tripolt:2017zgc,Strodthoff:2013cua}.
The effective average action
$\Gamma_k$ with a scale $k$ obeys the exact functional flow equation:
\begin{equation}
\displaystyle  \partial_k \Gamma_k=\frac{1}{2}STr\left[\frac{\partial_k R_k}{\Gamma^{(2)}_k+R_k}\right].
\end{equation}
Here, $\Gamma^{(2)}_k$ is the second functional derivative of the effective average action with respect to the fields; the trace includes momentum integration as well as traces of overall inner indices. An infrared regulator $R_k$ is introduced to suppress fluctuations at momenta below the scale $k$. Following the RG scale, the regulator may assume a functional form\cite{PAWLOWSKI2017165}.
In this investigation, quarks serve as the dynamical fields in the flow equation, $\sigma$, and $\boldsymbol{\pi}$, and they affect the effective potential and $\rho_{0}^3$ field. Contrary to the spatial components of the vector fields, the $\rho_{0}$ field is not dynamical because it is not coupled to the time derivative. Therefore, the value
of $\rho_{0}^3$ is completely fixed by specifying the values of other
fields. At each scale $k$ in the flow equation, we determine
the value of the $\rho_{0}^3$ field by solving the consistency equation
for given $\sigma$, and $\boldsymbol{\pi}$.
This $\rho^3_{0}$ field in turn appears in the effective
chemical potential for quarks, affecting the dynamical fluctuations in the flow equations. Throughout our study, we neglected the flow of all wave-function renormalization
factors in the so-called local potential approximation (LPA).

The scale-dependent effective potential can be expressed
by replacing the potential U with the scale-dependent
one $U_{k}$,
\begin{equation}
\displaystyle\Gamma_k =\int d^4 x{\mathcal{L}}|U\rightarrow U_k,
\end{equation}
with the Euclidean Lagrangian taken from Eq.\ref{ Lagrangian}, finite temperatures are treated
within the Matsubara formalism. The time-component is
Wick-rotated, ${t}\rightarrow -i\tau$, and the imaginary time $\tau$ is compactified on a circle with radius $\beta=\frac{1}{T}$, where $T$ is the temperature, thus introducing  
$\int d^4 x \equiv \int^{1/T}_0 d x_0 \int_V d^3 x$. 
Owing to the chiral symmetry, the potential $U$ depends on $\sigma$ and $\pi$ only through the chiral invariant:
\begin{equation}
\displaystyle   \phi^2 = \sigma^2+\boldsymbol{\pi}^2,
\end{equation}
As mentioned, the vector field $\rho^3_0$ appears here only as mean fields. The complete k-dependence is in the effective potential  $U_k$. In analogy to the mean-field potential, the effective potential has a chirally symmetric piece, $U_k^\phi$, the explicit chiral symmetry breaking term, and the mass terms of the vector bosons:
\begin{equation}
\displaystyle   U_k=U^\phi_k+U^\rho_k.
\end{equation}
Starting with some ultraviolet (UV) potentials $U_{\Lambda}$ as our initial conditions, we integrated fluctuations and obtained the scale-dependent $U_k$.
The form of  $U_k^\phi$ is determined without assuming any specific forms, while for the potential of the $\rho$ field:
 \begin{equation}
\displaystyle   U^{\rho}_k= -\frac{m^2_\rho}{2} (\rho_{0,k}^3)^2.
\end{equation}
To use Wetterich’s equation, a regulator function, which that respects the interpolating limits of the effective average action, has to be chosen. We employ the so-called optimized or Litim regulator function~\cite{Litim:2001up}, for bosons and fermions, respectively given by:
\begin{eqnarray}
    R^B_k(p) &=& (k^2-\boldsymbol{p}^2)\theta (k^2-\boldsymbol{p}^2), \\
    R^F_k(p) &=& \left(
                    \begin{array}{cc}
                        0 & ip_i(\gamma^E_i)^T \\
                        ip_i\gamma^E_i & 0 \\
                    \end{array}
                \right) (\sqrt{\frac{k^2}{p^2}-1})\theta (k^2-\boldsymbol{p}^2),
\end{eqnarray}
where, because of the structure of the regulators, the dependence on three-momenta is eliminated, and only the integral over theta function remains.
The flow equation for the potential $U_k^\phi$ can be obtained as follows:
\begin{eqnarray}\label{flow equation}
\partial_k U^\phi_k(T,\mu) &=& \frac{k^4}{12\pi^2}\Bigg [\left\{ \frac{3[1+2n_B(E_\pi)]}{E_\pi}+\frac{[1+2n_B(E_\sigma)]}{E_\sigma}\right\}\\
&-& \nu_q\left\{\frac{1-n_F(E_q,\mu^+_{eff})-n_F(E_q,-\mu^{-}_{eff})}{E_q} \right\}\\
&-& \nu_q\left\{\frac{1-n_F(E_q,-\mu^+_{eff})-n_F(E_q,\mu^-_{eff})}{E_q} \right\} \Bigg ].
\end{eqnarray}
Here, $\nu_{q} = 2\,(\text{spin}) \times 2\,(\text{flavor}) \times 3\,(\text{color}) = 12$ is the quark degeneracy factor, and $E_{q} = \sqrt{p^{2} + m_{\text{eff}}^{2}}$ with the effective quark mass $m_{\text{eff}} = g_{s}\sigma$.
The effective energies are expressed as follows:
\begin{eqnarray}
    E_\pi=\sqrt{k^2+M^2_\pi},\\
    E_\sigma=\sqrt{k^2+M^2_\sigma},\\
    E_q=\sqrt{k^2+M^2_q},
\end{eqnarray}
for pion, sigma-meson, and quark, respectively. The scale-dependent particle masses are as follows:
\begin{eqnarray}
    M^2_q=g^2\phi^2,\\
    M^2_\pi=2U'_k(\phi^2),\\      
    M^2_\sigma=2U'_k(\phi^2)+4 \phi^2 U''_k(\phi^2),
\end{eqnarray}
and we also define $ U'_k=\frac{\partial U_k}{\partial \phi^2}$.

The effective chemical potential,
\begin{equation}
    \mu^\pm_{eff}=\mu\pm (\frac{\mu_{I}
}{2}+g_\rho \rho^3_{0,k}),\\
\end{equation}
depends also on the field  $\rho_{0,k}$, and depend on the scale $k$.
The extended occupation numbers simplify to the usual Fermi-Dirac distribution functions for boson and fermion occupation numbers:
\begin{equation}
n_B(E)=\frac{1}{e^{\beta E}-1}, \quad n_F(E,\mu)=\frac{1}{e^{\beta(E-\mu)}+1}.
\end{equation}
According to the flow equation for the effective potential, the $\rho$ field can be calculated self-consistently. Therefore, at each momentum scale $k$~\cite{Drews:2014spa} we solve the equation for $\rho^{3}_{0,k}$:
  \begin{equation}
\begin{array}{l}
\displaystyle \frac{\partial U_k}{\partial \rho^3_{0,k}}=0.\\
\end{array}
\end{equation}
The dependence on $\rho^{3}_{0,k}$ appears in the mass term and the fermion loop. The flow equation reads:
 \begin{equation}\label{flow equation for rho}
\begin{aligned} 
\partial_{k} \rho^{3}_{0, k} = -\frac{{g_{\rho}} k^{4}}{\pi^{2} m_{\rho}^{2} E_{q}} & \left\{\frac{\partial}{\partial \mu_{eff}^{+}} \left[ n_{F} \left(E_{q}, \mu_{eff}^{+}\right) + n_{F}\left(E_{q}, -\mu_{eff}^{+}\right) \right] \right. \\ 
& \left. + \frac{\partial}{\partial \mu_{eff}^{-}} \left[ n_{F}\left(E_{q}, \mu_{eff}^{-}\right) + n_{F}\left(E_{q}, -\mu_{eff}^{-}\right) \right] \right\}.
\end{aligned}
\end{equation} 
This equation constitutes our flow equations for the $\rho$ field as a function of $\phi$.
Note that the flow equations for $\rho^3_{0}$ can be solved for a
given $\phi$, independently of the potential $U^{k}_{\phi}$
(which only tells us where the minimum of $\phi$ is).
To understand the behavior of $\rho^3_{0}$, we
ignore the $k$ dependence in $\mu_{eff}$ for now, and carry out
the integration over $k$.
Finally, the initial conditions for the flow equations must
be set up. The {UV} scale $\Lambda$ should be sufficiently large in
order to take into account the relevant fluctuation effects
and small enough to render the description in terms of the
model degrees of freedom realistic~\cite{Drews:2013hha}. In our calculation
we follow the choice of Ref. ~\cite{Schaefer:2004en}, $\Lambda$ = 500 MeV. The initial condition for the potential is
 \begin{equation}\label{lambda}
 \displaystyle U^{\phi}_{\Lambda} = \frac{\lambda}{4}{\phi^4},
 \end{equation}
and the parameters are set as $g_{s}$ = 3.2, $\lambda$ = 8 with the vacuum
effective potential from the FRG computation having the
minimum at $\sigma_{vac}\simeq 93 $ MeV, which is regarded as $f_{\pi}$. We
set the value of $\lambda$, which enforces $\phi$ to stay near $f_{\pi}$, as $(\lambda \sim 8)$.\ If another initial condition with an additional $\phi^{2}$
term is employed to give the mass, $\lambda$ must be adjusted to obtain
qualitatively similar results; in fact, starting with the
condition described by Eq.(~\ref{lambda}), the scale evolution first generates the
$\phi^{2}$ terms, reflecting the universality.
The initial condition for the $\rho$ field was not been
examined in detail, and we simply set
\begin{equation}
    \begin{array}{cc}
     \displaystyle \rho^3_{0,\Lambda}(\phi) = 0.
    \end{array}
\end{equation}
We also set a different initial condition but do not present it here, as it does not affect our main results.

Assembling all these elements, we calculated the effective
potential with the fluctuations integrated for $\Gamma_{IR}$ = 0. The
final step is to find $\Phi$ = $\sigma^{*}$, which minimizes the effective
potential. At the location of minimum, the effective potential is identified as the thermodynamic potential,
 \begin{equation}
     \begin{array}{l}
          \displaystyle \frac{T\Omega(\mu,T)}{V} = \Gamma_{IR=0}(\mu,T,\sigma^{*}),
        
     \end{array}
 \end{equation}
 In practice, it is numerically expensive to reduce the $IR$
cutoff, and we typically stopped the integration under
$k_{IR} \simeq20 MeV$.

\section {RESULTS}\label{sec:3}

\subsection{Chiral Phase Diagram with isospin couplings}

In this section, we analyze the effects of varying isospin chemical potentials and coupling constants on the chiral phase transition in the chiral limit. The analysis is performed using the FRG flow equation for the $\rho$ meson, given in Eq.~(\ref{flow equation for rho}), over a range of temperatures and chemical potentials. Additionally, we determine the $\rho$ meson condensate by solving the FRG effective potential equation, that is, Eq.~(\ref{flow equation for rho}), for different parameter sets to facilitate comparison.
\begin{figure}
    \centering
\includegraphics[width=0.7\textwidth]{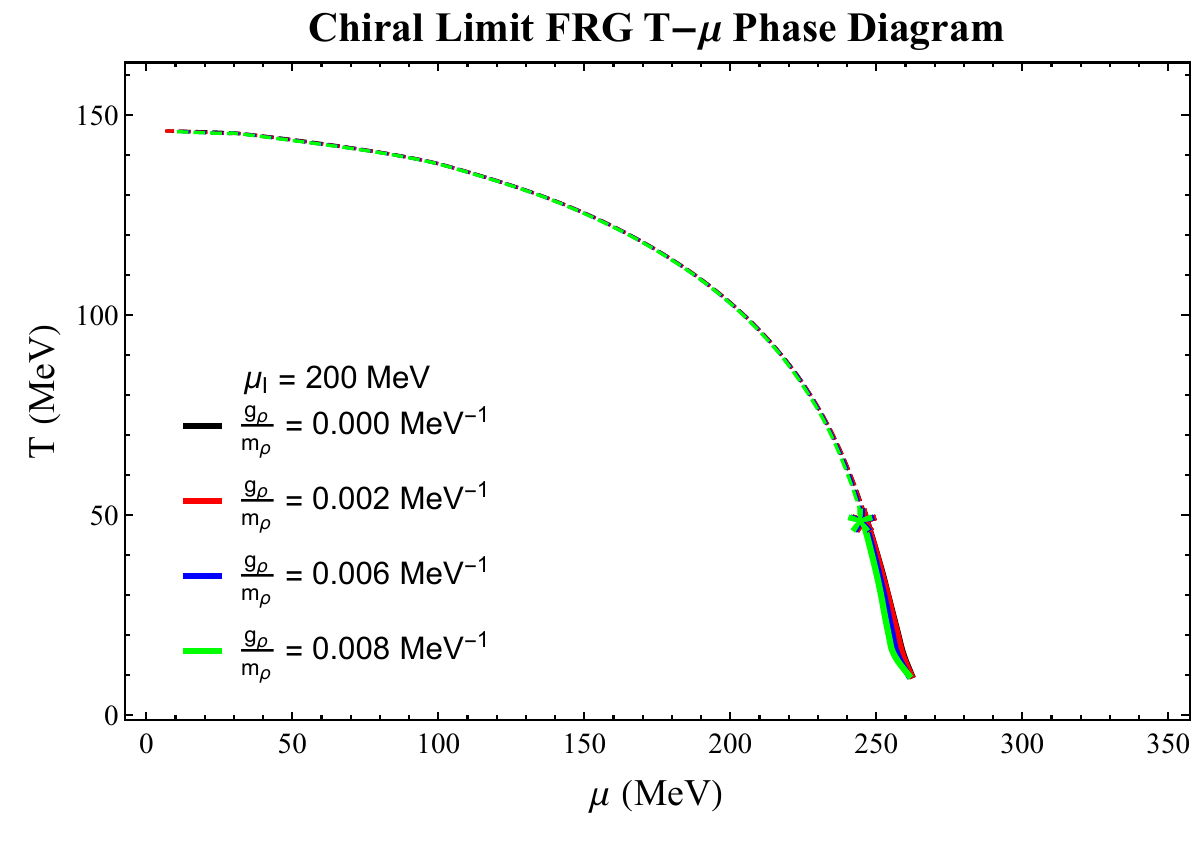}
    \caption{ (color online) FRG ${T-\mu}$ chiral phase diagram with different vector couplings. The solid lines show the first-order phase transition, whereas the dashed lines show the second-order phase transition. The stars show the TCPs. The parameters were set as $f_{\pi}$ = 93 MeV, $g_{s}$ = 3.2, $\lambda$ = 8, the ultraviolet cutoff   $\Lambda_{FRG}$ =500 MeV, and ${\mu_I}$ = 200 MeV.}
    \label{fig:The FRG rho meson phase diagram with different vector couplings}
\end{figure}
  
\begin{figure}
    \centering
    \includegraphics[width=0.7\textwidth]{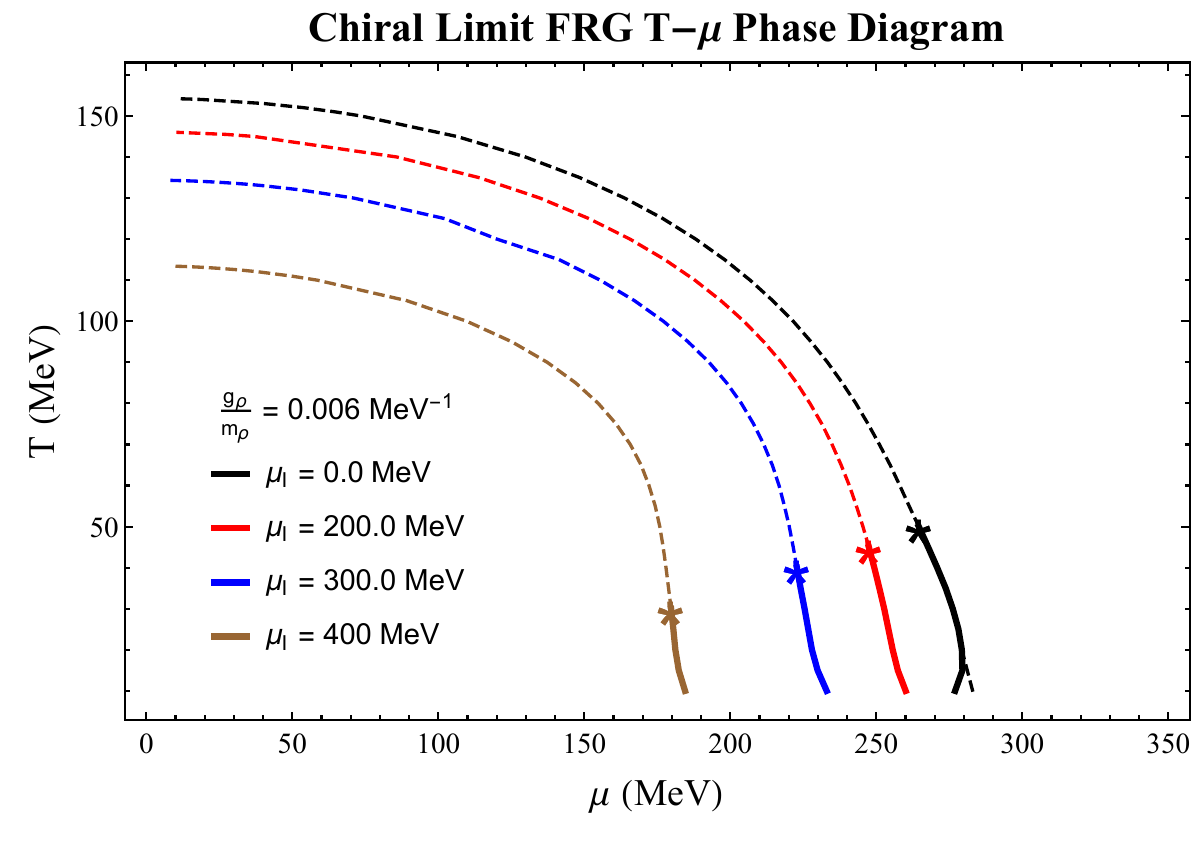}
    \caption{(color online) The FRG ${T-\mu}$ chiral phase diagram with different $\mu_{I}$. The solid lines represent the first-order phase transition, and the dashed lines represent the second-order phase transition. The stars show the TCPs. The parameters are set as: $f_{\pi}$ = 93 MeV, $g_{s}$ = 3.2, $\lambda$ = 8, the ultraviolet cutoff   $\Lambda_{FRG}$ =500 MeV, the coupling constant ${g_{\rho}}$ $m^{-1}_{\rho} = 0.006 \ \mathrm{MeV}^{-1}$.}
   \label{fig:The FRG phase diagram with different isospin chemical potential}
\end{figure}

Fig.~\ref{fig:The FRG rho meson phase diagram with different vector couplings} shows the impact of the vector coupling constants on the chiral phase boundary. Note that, at fixed temperatures, the chiral phase transition occurs at higher chemical potentials. Although $\rho$ vector mesons contribute to the increased stability of the system, their influence on the phase boundary becomes progressively limited as the coupling strength increases. The boundary of the first-order phase transition shifts to lower temperatures.

\begin{figure}
\centering
\includegraphics[width=0.7\textwidth]{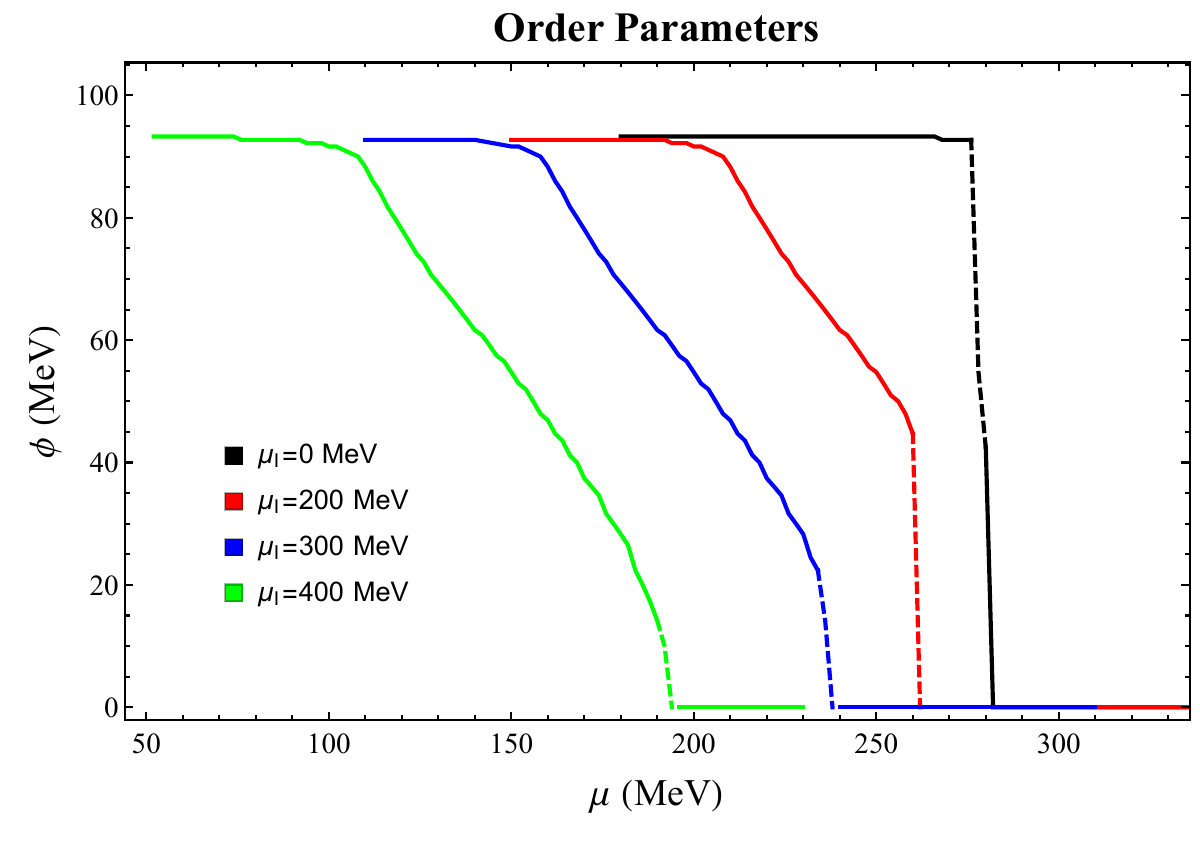}
\caption{\label{fig: chiral condensate diff isosspin(10)} (color online) Chiral condensates as a function of quark chemical potential under different isospin chemical potentials, calculated at $T=10$ MeV, ${g_{\rho}}/m_{\rho} = 0.006~\mathrm{MeV}^{-1}$. The different colored lines correspond to different isospin chemical potentials.}
\end{figure}

The chiral phase diagram derived from the functional renormalization group equation at various isospin chemical potentials is presented in Fig.~\ref{fig:The FRG phase diagram with different isospin chemical potential}. As the isospin chemical potential increases, the chemical potential required for the chiral phase transition decreases, leading to a leftward shift of the critical endpoint toward lower temperature and chemical potential regions, where the RG flow evolution is significantly influenced by low-temperature fluctuations. This shift of the chiral boundary with increasing $\mu_I$ is consistent with both analytical calculations and lattice QCD results\cite{Brauner:2016lkh}. While Ref.~\cite{osman2025functional}reported comparable behavior under specific conditions, the present study extends these findings and provides a more comprehensive demonstration.

To accurately understand the process of phase transition under the influence of isospin chemical potential, see Fig.~\ref{fig: chiral condensate diff isosspin(10)}. The change in order parameters in the low-temperature region of the phase diagram with chemical potential can be observed. Evidently, a first-order phase transition is occurring in this region. As the chemical potential increases, the chiral condensate changes and drops to zero at the phase transition. Note also that the initial value of the chiral condensate gradually decreases with the increase of the isospin chemical potential. It is hypothesized that when the isospin chemical potential gradually increases and reaches a certain value, the first-order phase transition region eventually disappears; this is also shown phase diagram.  
\subsection{$\rho$ meson condensate}
Based on the analyses of Figs.~\ref{fig:The FRG rho meson phase diagram with different vector couplings} and \ref{fig:The FRG phase diagram with different isospin chemical potential}, we selected optimal parameter values to compute the $\rho$ meson condensation, considering different couplings ${g}_{\rho}/m_{\rho}$. These choices were guided by the observed shifts in the phase boundary and the influence of varying isospin chemical potentials on the stability of the system. The data for ${g}_{\rho} \rho^{3}_{0}$ were obtained by directly solving Eq.~\ref{flow equation for rho}. The figures illustrate how combinations of temperature, chemical potential, and coupling strength affect meson condensation. By carefully tuning these variables, we accurately determined the condensation values, providing deeper insights into chiral phase transitions under varying conditions.
Near chiral restoration, vector and axial‑vector modes broaden and nearly degenerate. The $\rho$ channel competes with the chiral condensate and reshapes $U_k$, producing the mutual ``pull'' observed in the curves. This reflects a coupled dynamical interplay rather than two separable phenomena, arising from their shared origin in the QCD Lagrangian, where both structures emerge from the same fermionic degrees of freedom and their interactions. These results, which highlight the role of the $\rho$ meson in stabilizing the system, particularly in regions of the phase diagram where fluctuations are pronounced, and are essential for understanding the behavior of hadronic matter in extreme environments such as heavy-ion collisions or astrophysical settings\cite{STAR:2005gfr,Harada:2005br}.
\begin{figure}[H]
     \centering
     \begin{subfigure}[b]{0.45\textwidth}
         \centering
         \includegraphics[width=\textwidth]{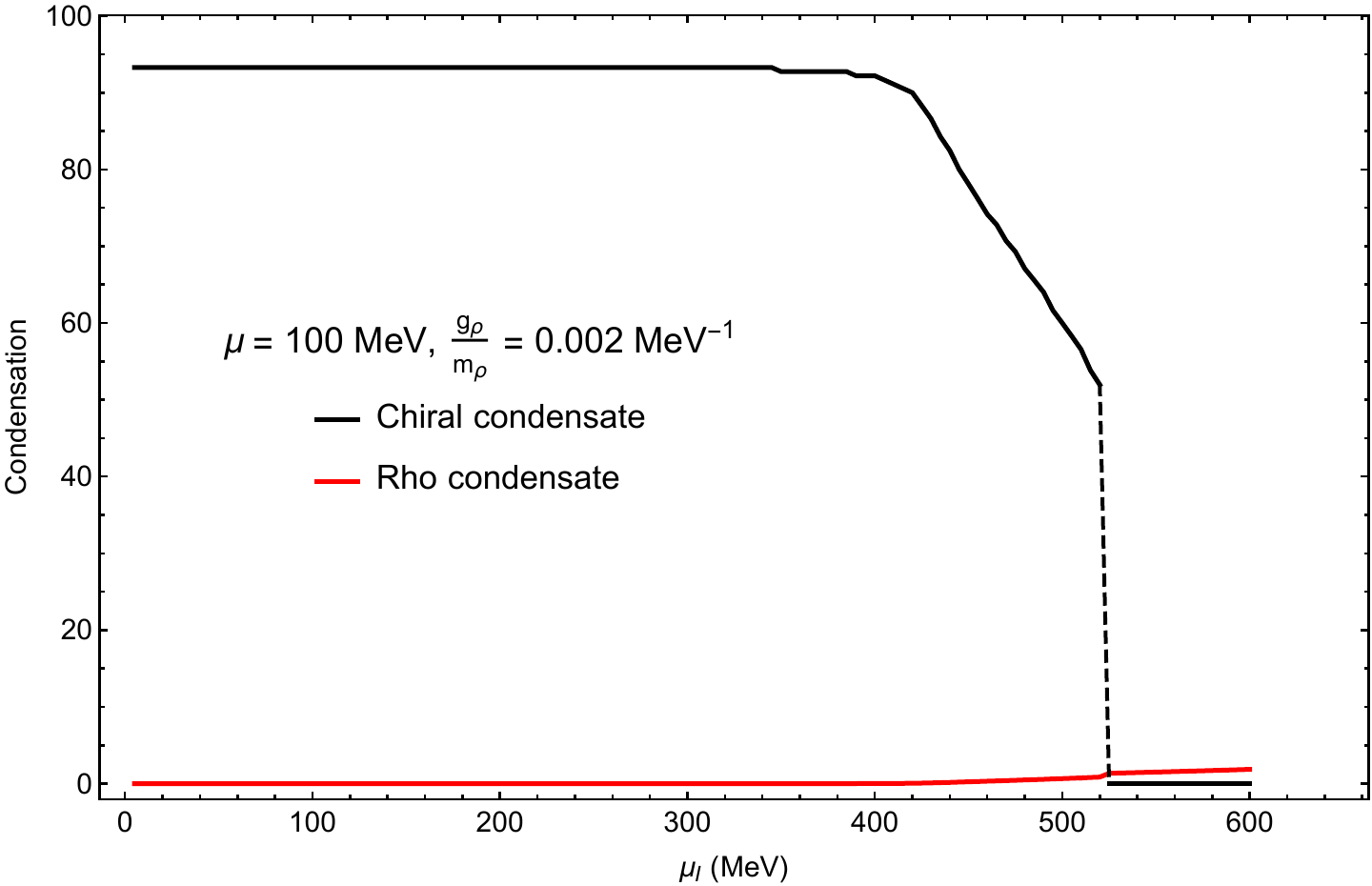}
         \caption{}
         \label{}
     \end{subfigure}
     \hfill
     \begin{subfigure}[b]{0.45\textwidth}
         \centering
         \includegraphics[width=\textwidth]{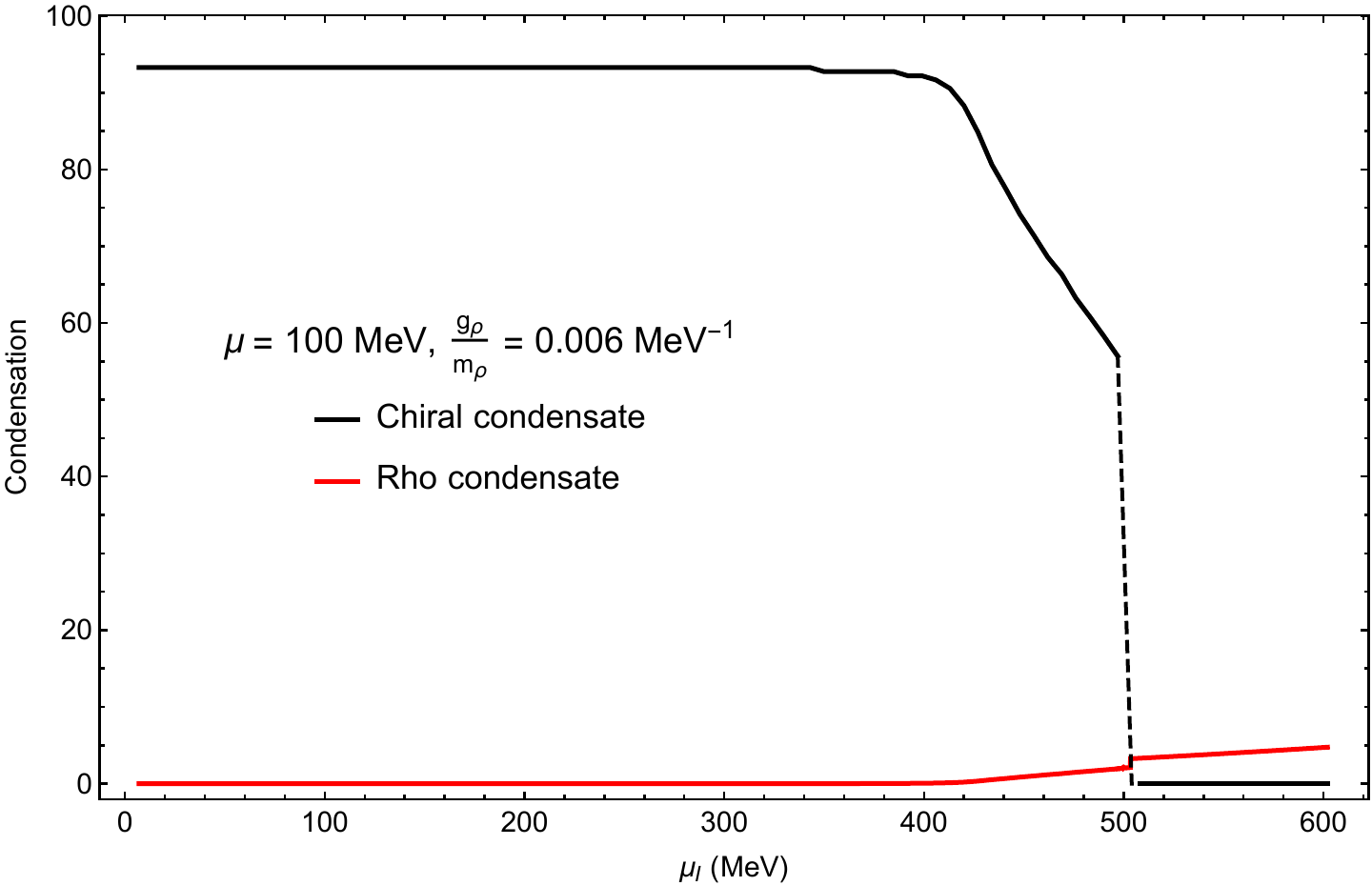}
         \caption{}
         \label{}
     \end{subfigure}
     \hfill
     \begin{subfigure}[b]{0.45\textwidth}
         \centering
         \includegraphics[width=\textwidth]{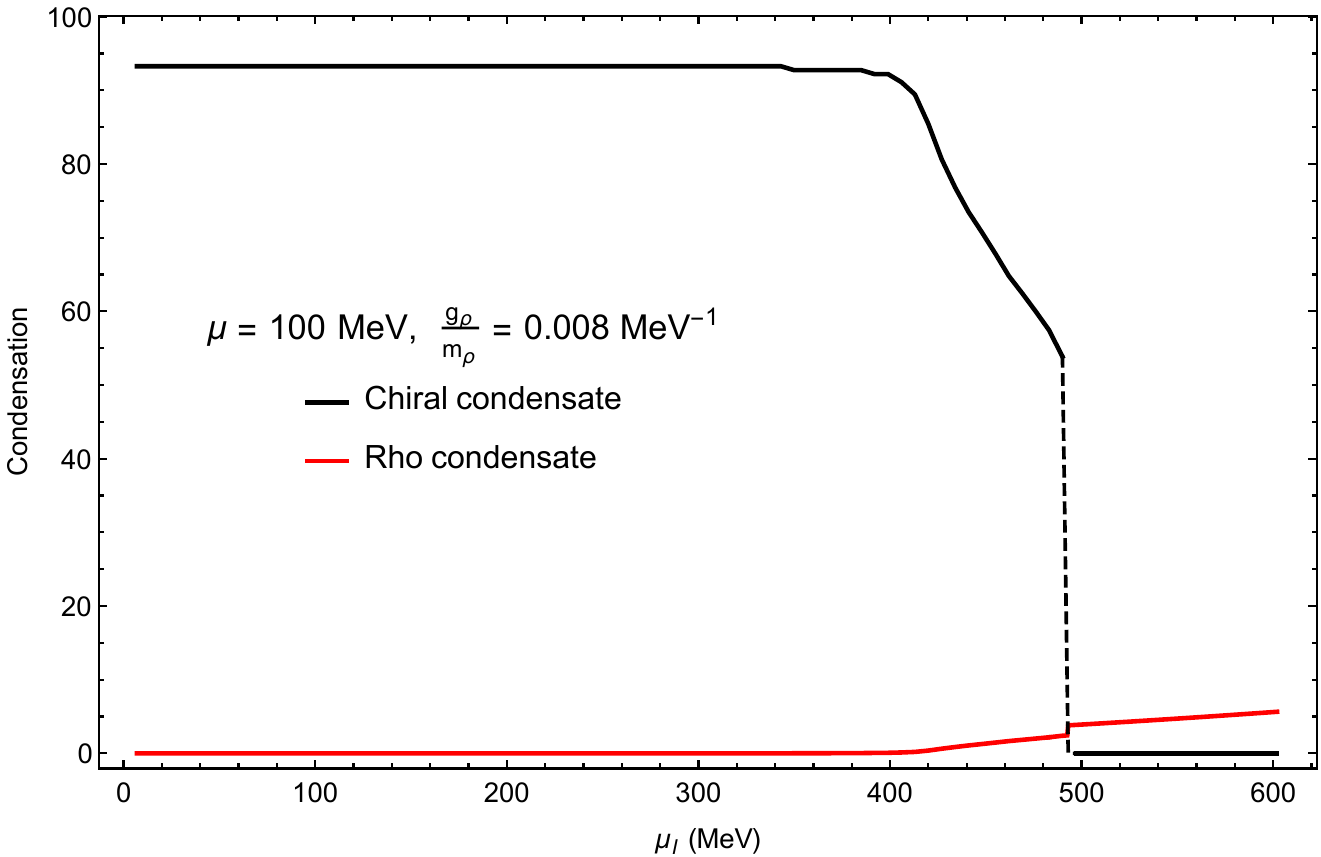}
         \caption{}
         \label{}
     \end{subfigure}
      \hfill
     \begin{subfigure}[b]{0.45\textwidth}
         \centering
         \includegraphics[width=\textwidth]{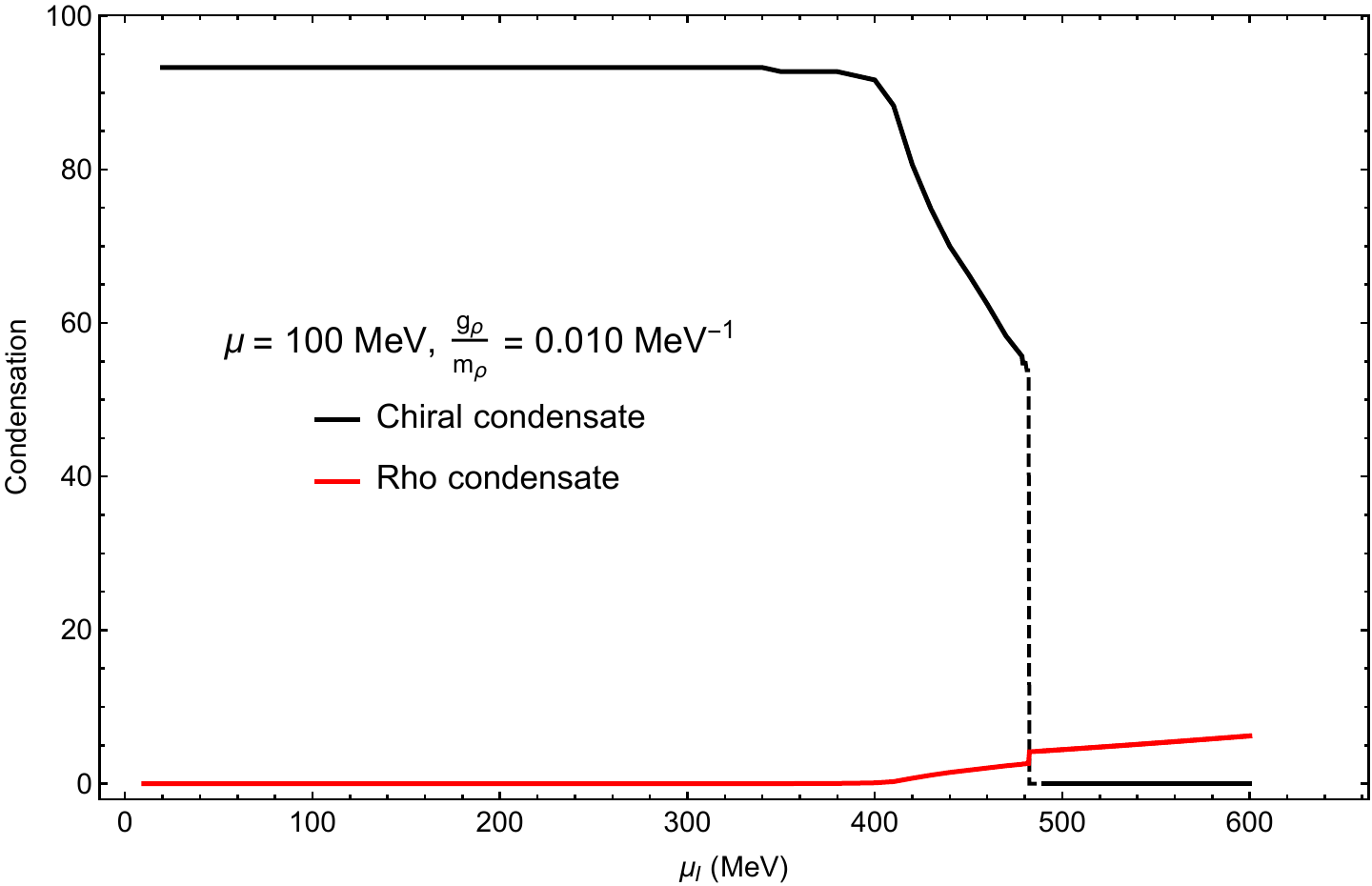}
         \caption{}
         \label{}
     \end{subfigure}
        \caption{(color online) Chiral and $\rho$ condensates as a functions of the isospin chemical potential: (a) $\frac{{g_{\rho}}}{m_{\rho}} = 0.002$ MeV$^{-1}$, (b) $\frac{{g_{\rho}}}{m_{\rho}} = 0.006$ MeV$^{-1}$, (c) $\frac{{g_{\rho}}}{m_{\rho}} = 0.008$ MeV$^{-1}$, and (d) $\frac{{g_{\rho}}}{m_{\rho}} = 0.010$ MeV$^{-1}$. The chemical potential was set as $\mu = 100$ MeV, temperature was set as $T = 10$ MeV and the ultraviolet cutoff was set as $\Lambda_{FRG}$ =500 MeV}
        \label{The chiral and rho condensates as a function of isospin chemical (1)}
\end{figure}
In Fig.~\ref{The chiral and rho condensates as a function of isospin chemical (1)}, we compare the results for the $\rho$ condensate as a function of isospin chemical potential, for $\mu$ = $100$ MeV, and different $\rho$ couplings: panel (a): ${g_{\rho}}$ $m^{-1}_{\rho}$= 0.002 MeV$^{-1}$,  panel (b): ${g_{\rho}}$ $m^{-1}_{\rho}$= 0.006 MeV$^{-1}$, panel (c): ${g_{\rho}}$ $m^{-1}_{\rho}$= 0.008 MeV$^{-1}$, and panel (d): ${g_{\rho}}$ $m^{-1}_{\rho}$= 0.010 MeV$^{-1}$. The temperature for these calculations was set as T = 10 MeV.
In this scenario, the chiral condensate decreases with increasing $\mu_{I}$ while the $\rho$ condensate starts to grow for $\mu_{I}$ greater than the critical value. At even higher isospin chemical potential, a new ﬁrst-order transition occurs, and the chiral condensate drops to zero. In this new region, the $\rho$ condensate becomes dominant when the coupling value for the $\rho$ meson increases.

\begin{figure}[H]
     \centering
     \begin{subfigure}[b]{0.45\textwidth}
         \centering
         \includegraphics[width=\textwidth]{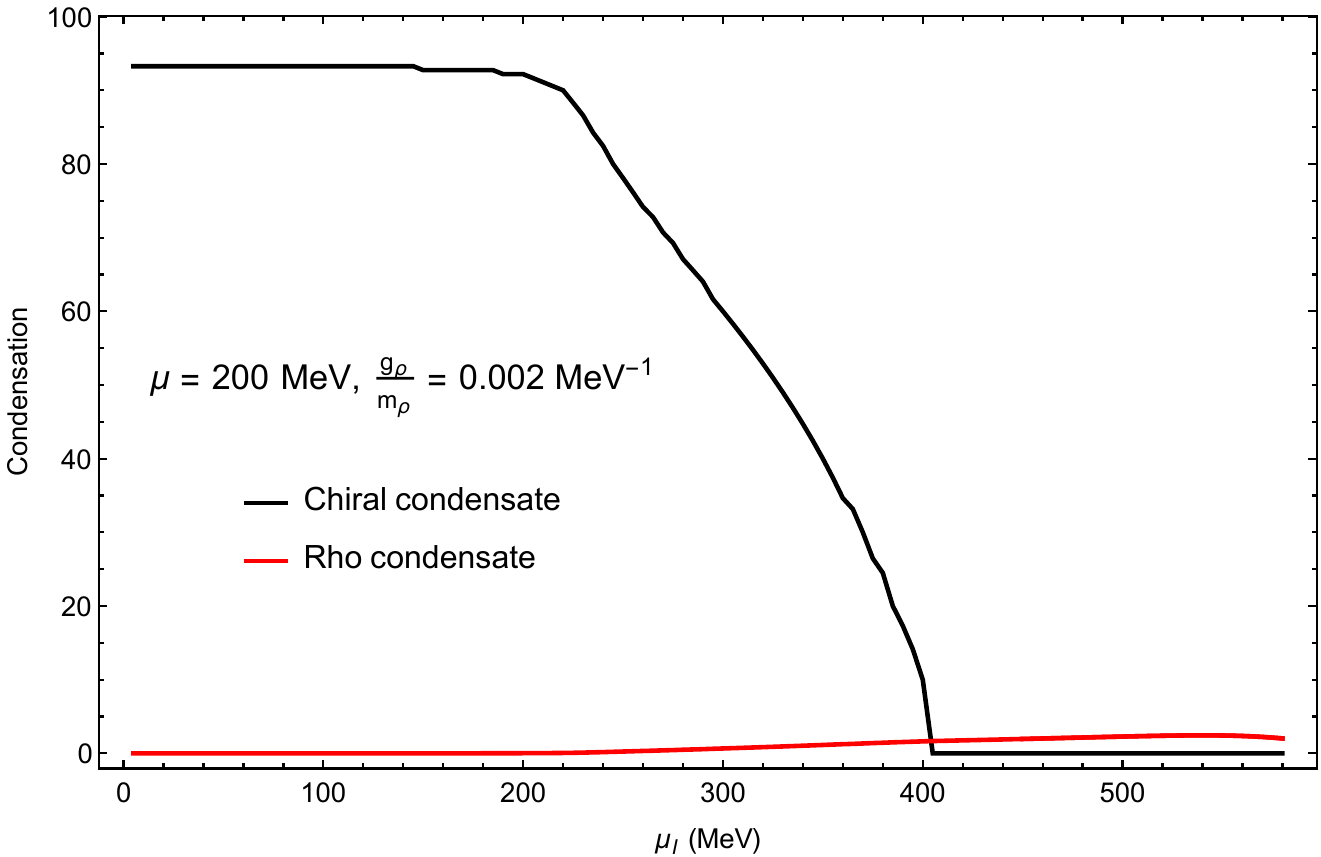}
         \caption{}
         \label{}
     \end{subfigure}
     \hfill
     \begin{subfigure}[b]{0.45\textwidth}
         \centering
         \includegraphics[width=\textwidth]{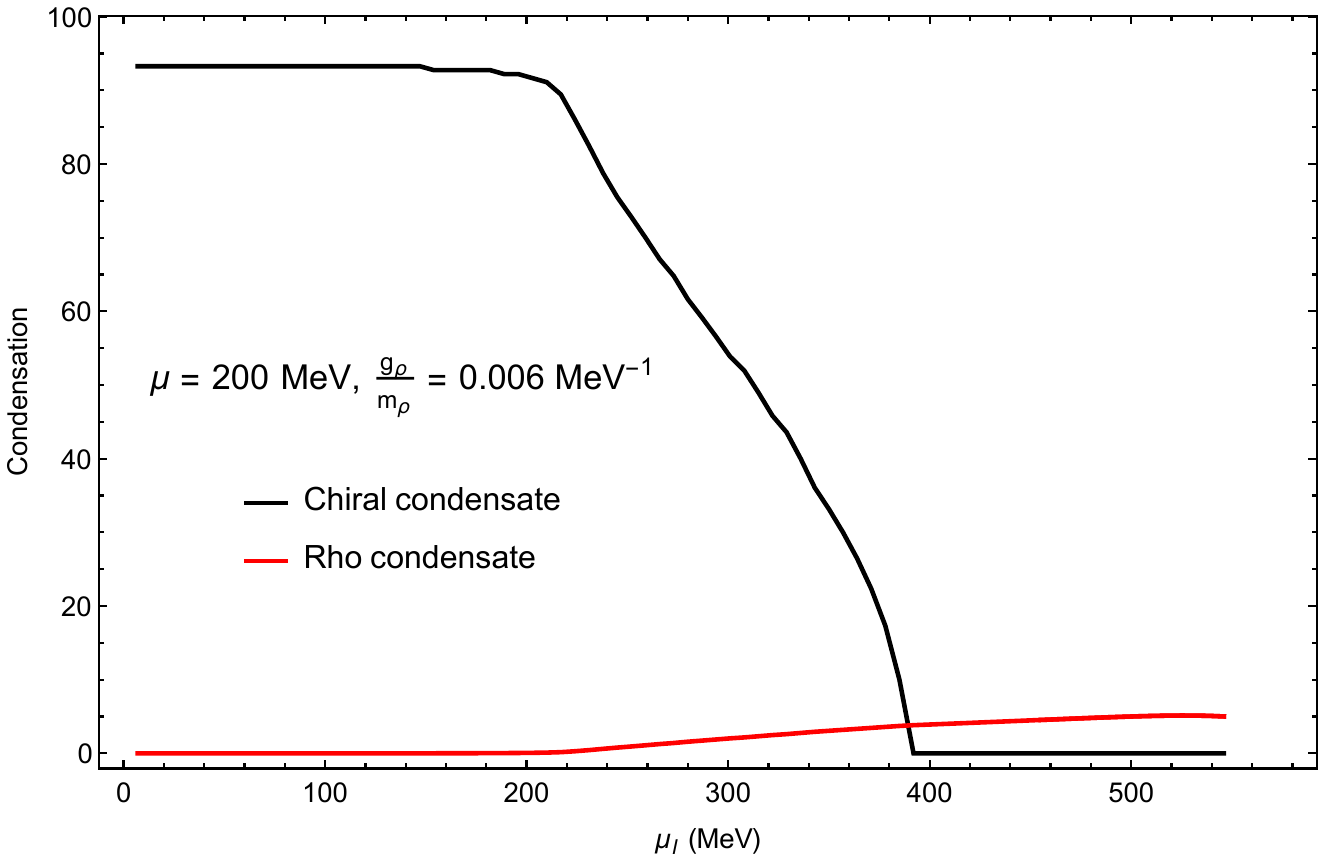}
         \caption{}
         \label{}
     \end{subfigure}
     \hfill
     \begin{subfigure}[b]{0.45\textwidth}
         \centering
         \includegraphics[width=\textwidth]{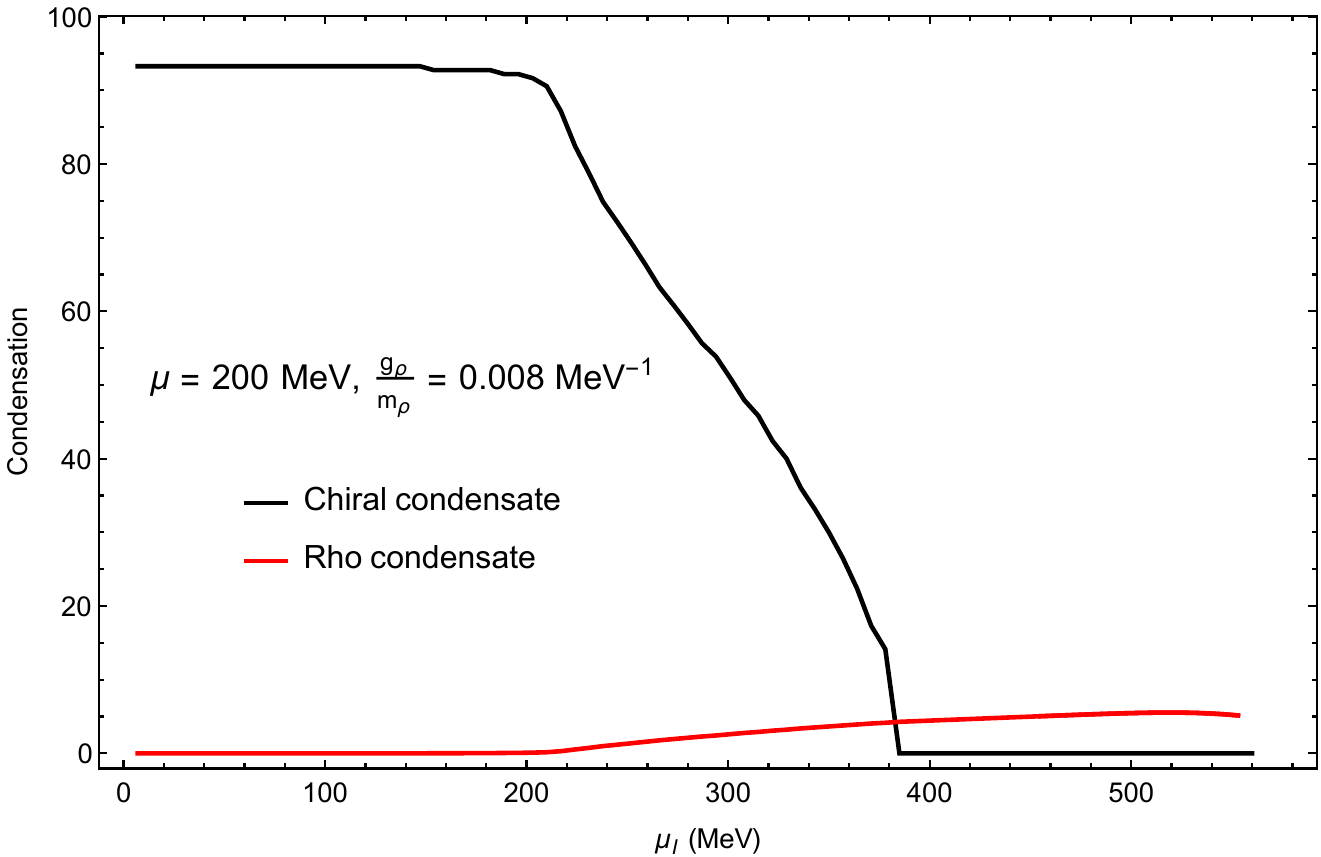}
         \caption{}
         \label{}
     \end{subfigure}
      \hfill
     \begin{subfigure}[b]{0.45\textwidth}
         \centering
         \includegraphics[width=\textwidth]{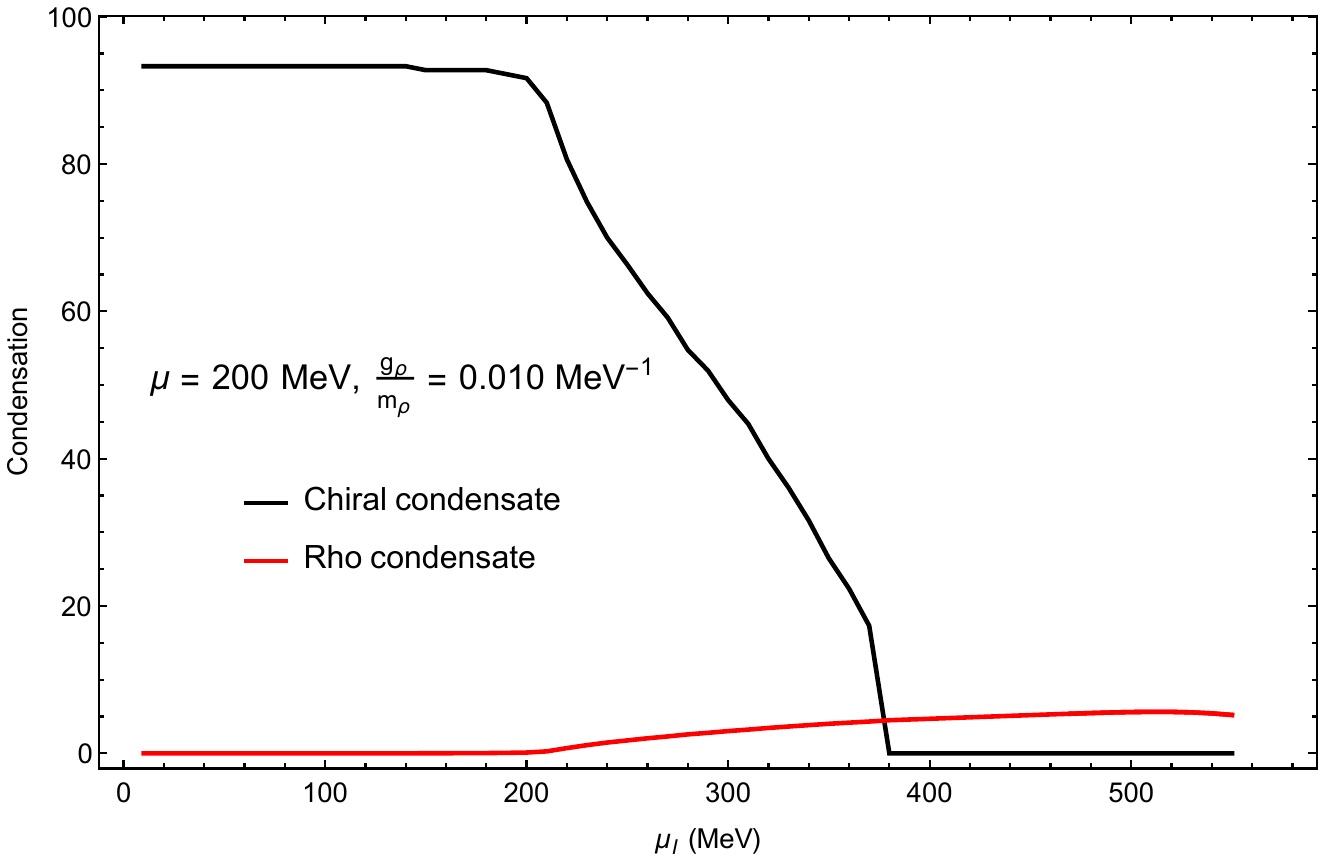}
         \caption{}
         \label{}
     \end{subfigure}
        \caption{(color online) Chiral and $\rho$ condensates $\rho$ as function of the isospin chemical potential: (a) $\frac{{g_{\rho}}}{m_{\rho}} = 0.002$ MeV$^{-1}$, (b) $\frac{{g_{\rho}}}{m_{\rho}} = 0.006$ MeV$^{-1}$, (c) $\frac{{g_{\rho}}}{m_{\rho}} = 0.008$ MeV$^{-1}$, and (d) $\frac{{g_{\rho}}}{m_{\rho}} = 0.010$ MeV$^{-1}$. The chemical potential was set as $\mu = 200$ MeV, the temperature was set as $T = 10$ MeV, and the ultraviolet cutoff was set as $\Lambda_{FRG}$ =500 MeV}
        \label{The chiral and rho condensates as a function of isospin chemical (2)}
\end{figure}

\begin{figure}[H]
     \centering
     \begin{subfigure}[b]{0.45\textwidth}
         \centering
         \includegraphics[width=\textwidth]{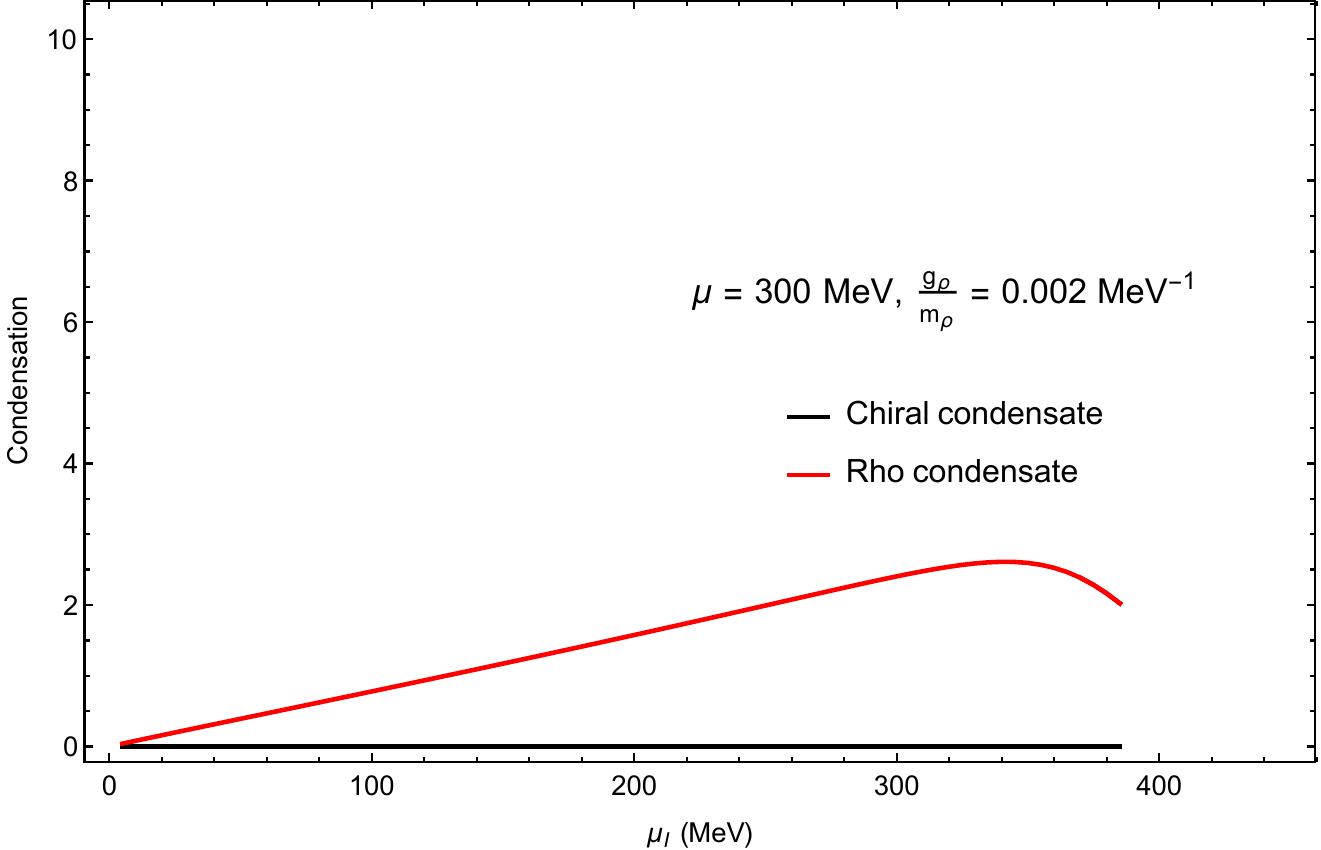}
         \caption{}
         \label{}
     \end{subfigure}
     \hfill
     \begin{subfigure}[b]{0.45\textwidth}
         \centering
         \includegraphics[width=\textwidth]{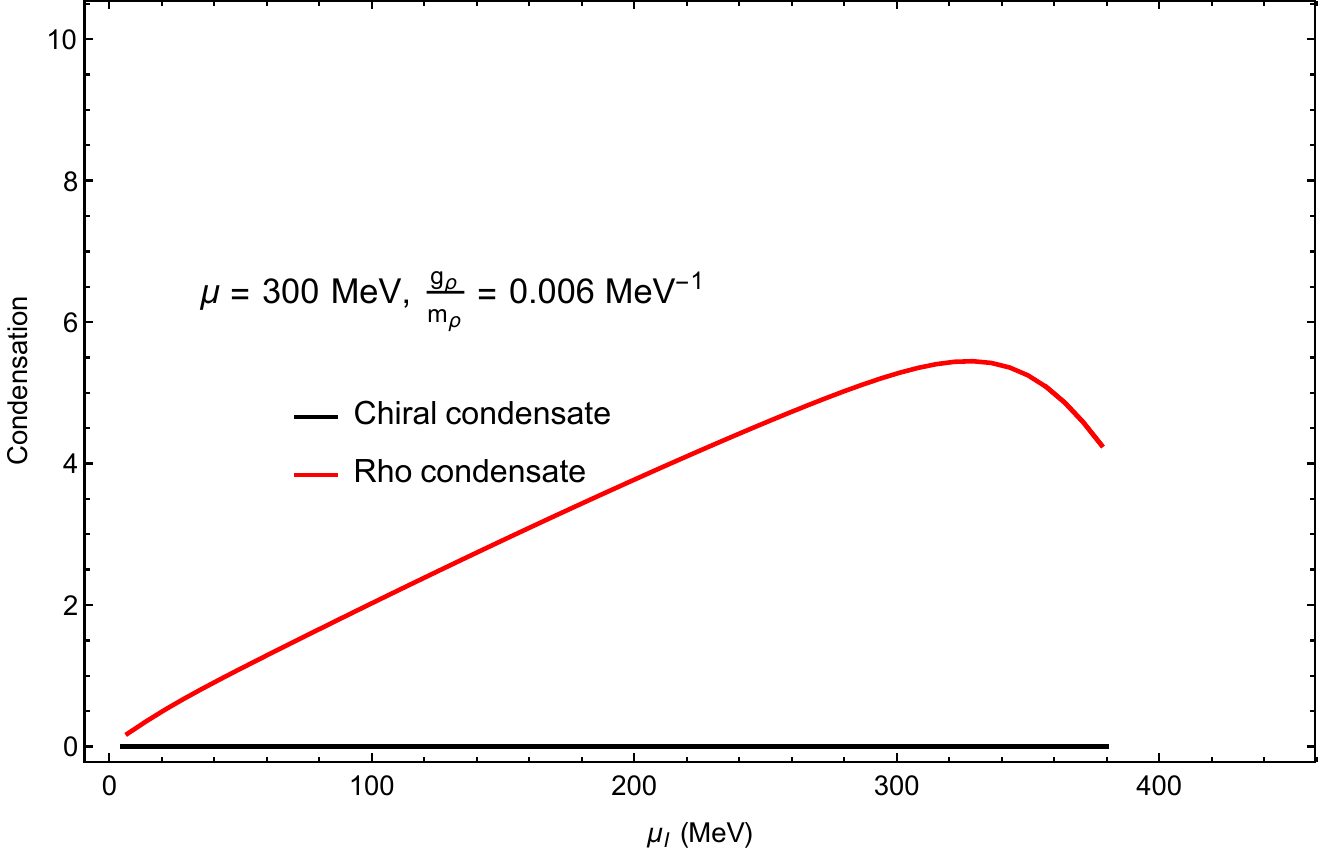}
         \caption{}
         \label{}
     \end{subfigure}
     \hfill
     \begin{subfigure}[b]{0.45\textwidth}
         \centering
         \includegraphics[width=\textwidth]{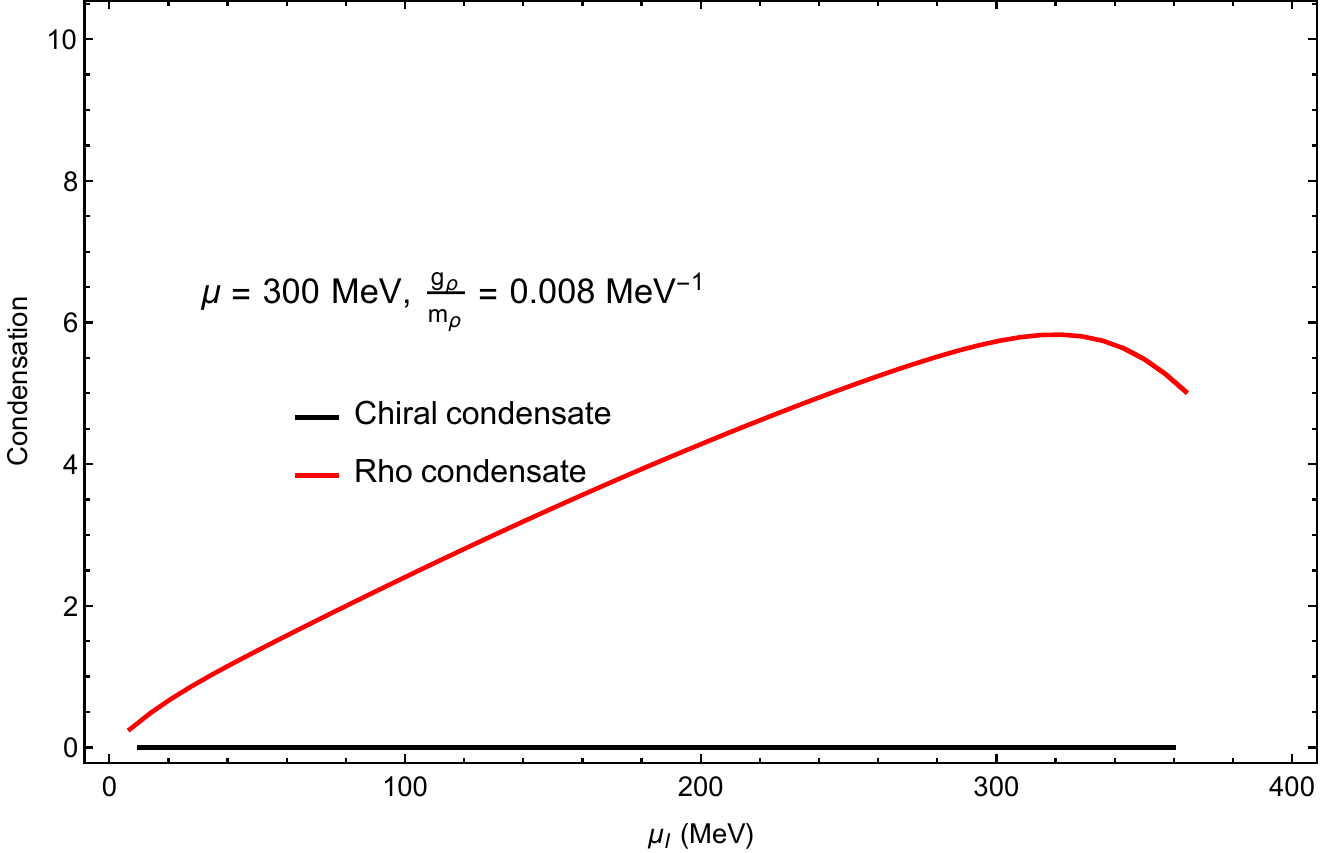}
         \caption{}
         \label{}
     \end{subfigure}
      \hfill
     \begin{subfigure}[b]{0.45\textwidth}
         \centering
         \includegraphics[width=\textwidth]{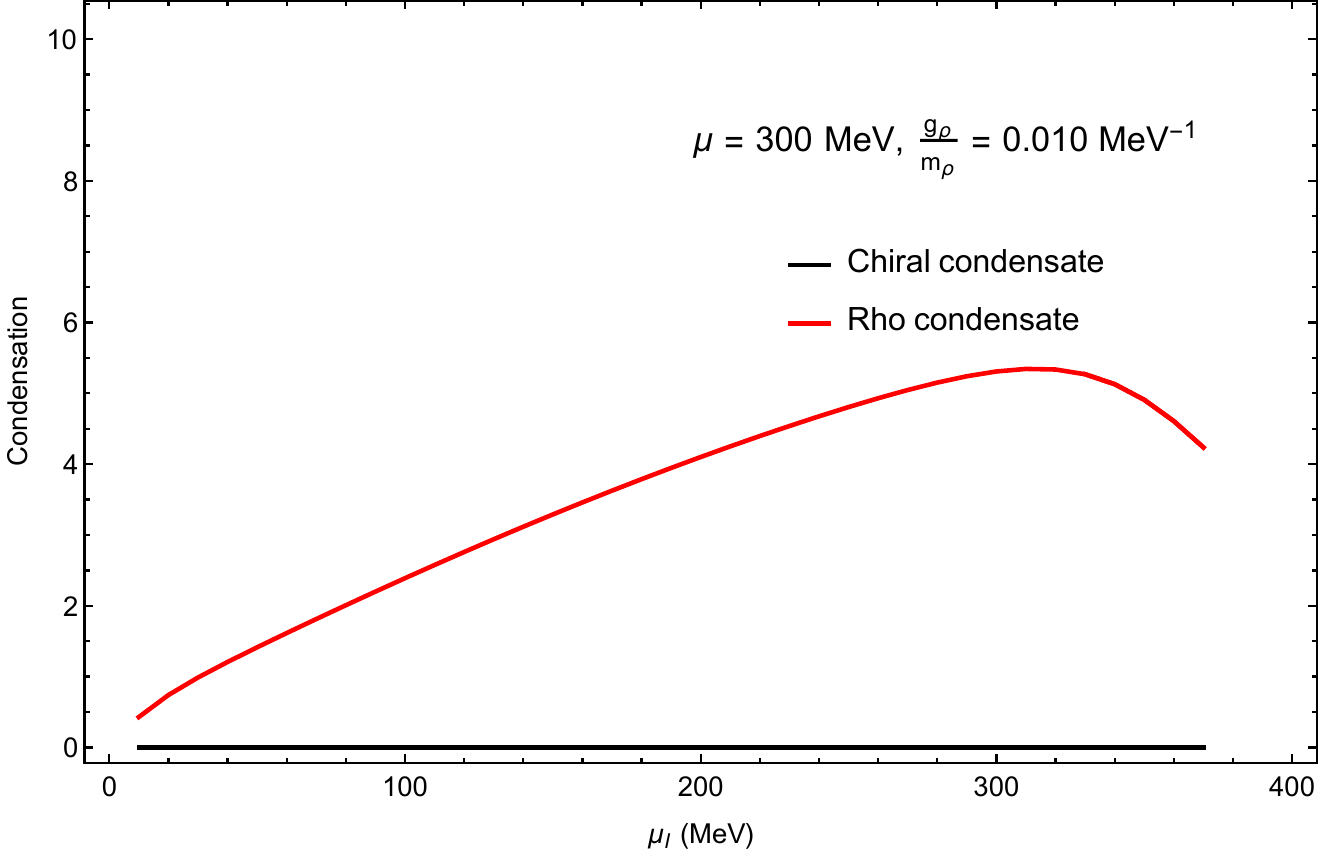}
         \caption{}
         \label{}
     \end{subfigure}
        \caption{Chiral and $\rho$ condensates $\rho$ as a functions of the isospin chemical potential: (a) $\frac{{g_{\rho}}}{m_{\rho}} = 0.002$ MeV$^{-1}$, (b) $\frac{{g_{\rho}}}{m_{\rho}} = 0.006$ MeV$^{-1}$, (c) $\frac{{g_{\rho}}}{m_{\rho}} = 0.008$ MeV$^{-1}$, and (d) $\frac{{g_{\rho}}}{m_{\rho}} = 0.010$ MeV$^{-1}$. The chemical potential was set as $\mu = 300$ MeV, the temperature was set as $T = 10$ MeV, and the ultraviolet cutoff was set as $\Lambda_{FRG}$ =500 MeV}
        \label{The chiral and rho condensates as a function of isospin chemical (4)}
\end{figure}

Fig.~\ref{The chiral and rho condensates as a function of isospin chemical (2)} shows the chiral and $\rho$ condensates as functions of the isospin chemical potential for $\mu = 200$ MeV. Across all panels, the behavior differs from that at lower $\mu$. The $\rho$ condensate begins to grow when $\mu_I$ exceeds the critical value for a second-order phase transition, while the chiral condensate simultaneously drops to zero. In this regime, the $\rho$ condensate remains dominant for larger $\rho$ meson couplings.
The chiral condensate vanishes in all panels of Fig.~\ref{The chiral and rho condensates as a function of isospin chemical (4)}, at $T = 10~\mathrm{MeV}$ and $\mu = 300~\mathrm{MeV}$. Comparing $g_\rho / m_\rho = [0.002,\, 0.010]~\mathrm{MeV}^{-1}$, one finds that larger $g_\rho$ values enhance isovector repulsion, causing the $\rho$ condensate to dominate the order-parameter dynamics while the chiral condensate becomes almost completely suppressed. At strong coupling, the system is increasingly governed by the vector channel, reinforcing the conclusion that isovector repulsion reshapes the low-temperature phase structure\cite{Tripolt:2017zgc}.
In Figs. ~\ref{The chiral and rho condensates as a function of isospin chemical (1)},~\ref{The chiral and rho condensates as a function of isospin chemical (2)} and ~\ref{The chiral and rho condensates as a function of isospin chemical (4)}, we observe the influence of the chemical potential on the chiral and $\rho$ condensates; the isospin chemical potential takes different critical values when the transition of the $\rho$ condensate occurred from zero to non-zero.

\begin{figure}[H]
\centering
\begin{subfigure}[b]{0.45\textwidth}
\includegraphics[width=\textwidth]{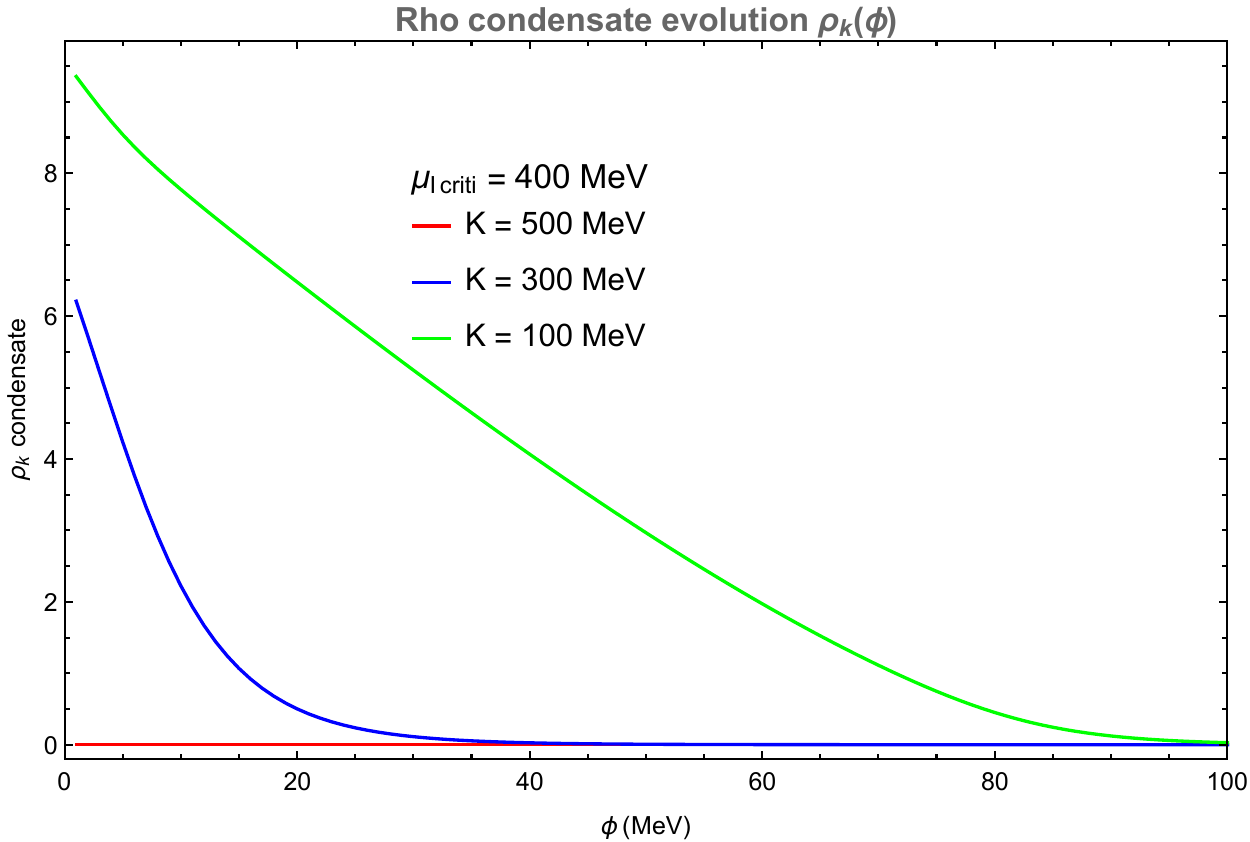} 
\caption{}
\label{}
\end{subfigure}
\hfill
\begin{subfigure}[b]{0.45\textwidth}
\includegraphics[width=\textwidth]{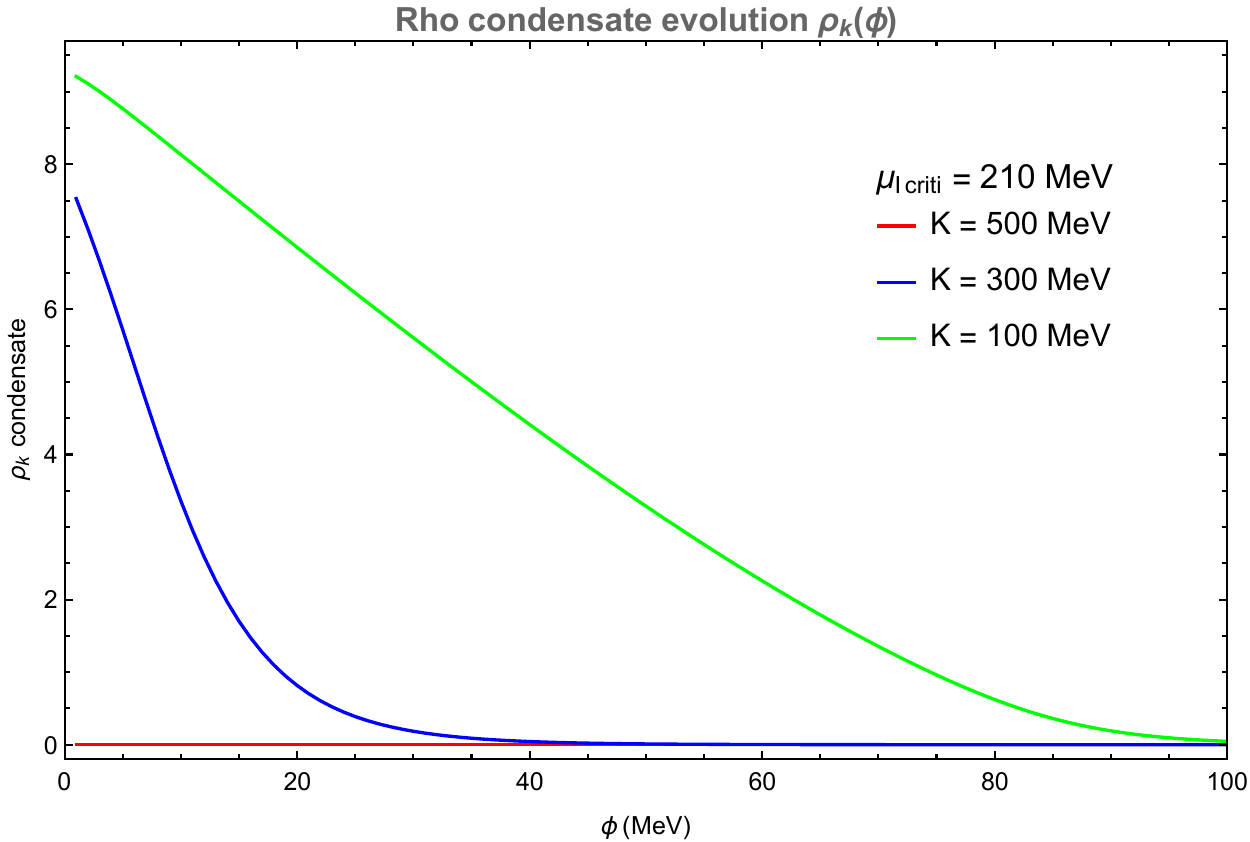}
\caption{}
\label{}
\end{subfigure}
\vskip\baselineskip
\begin{subfigure}[b]{0.45\textwidth}
\includegraphics[width=\textwidth]{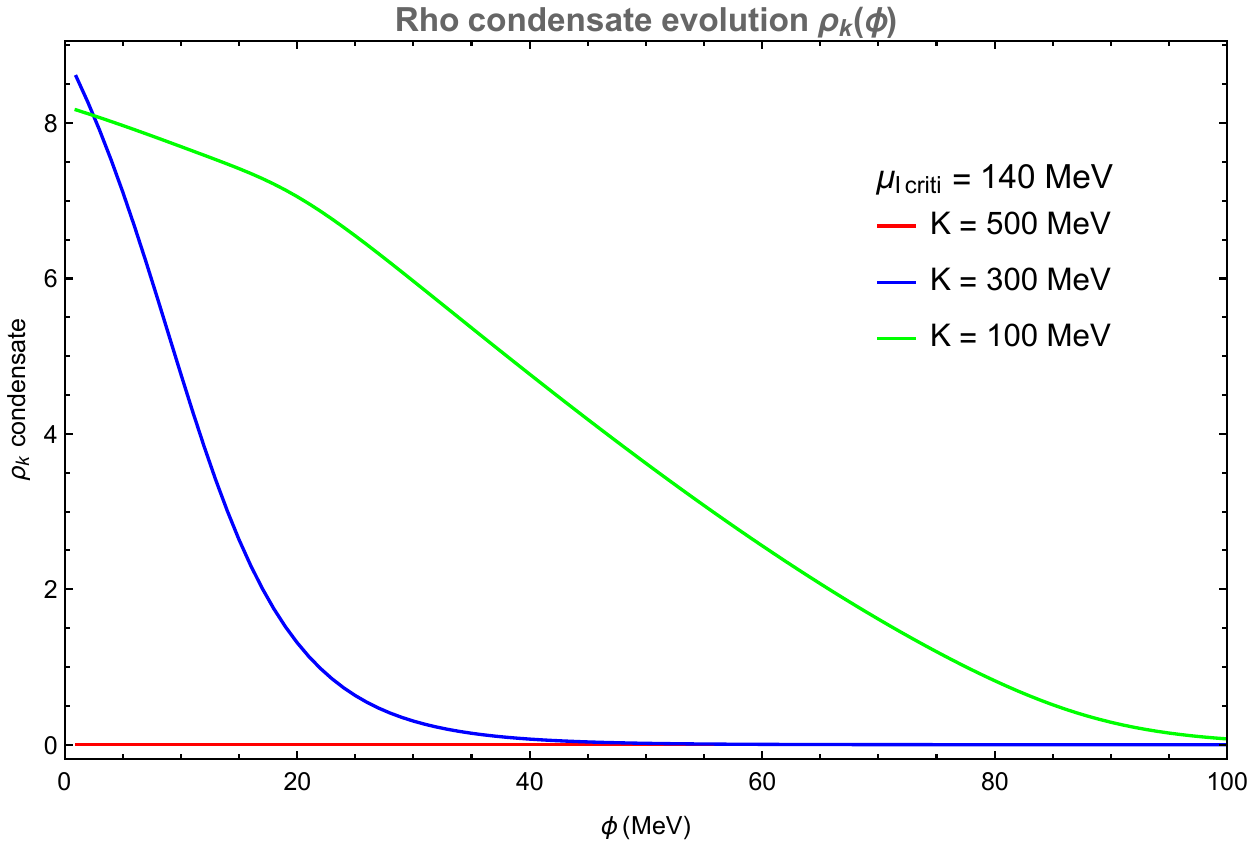} 
\caption{}
\label{}
\end{subfigure}
\hfill
\begin{subfigure}[b]{0.45\textwidth}
\includegraphics[width=\textwidth]{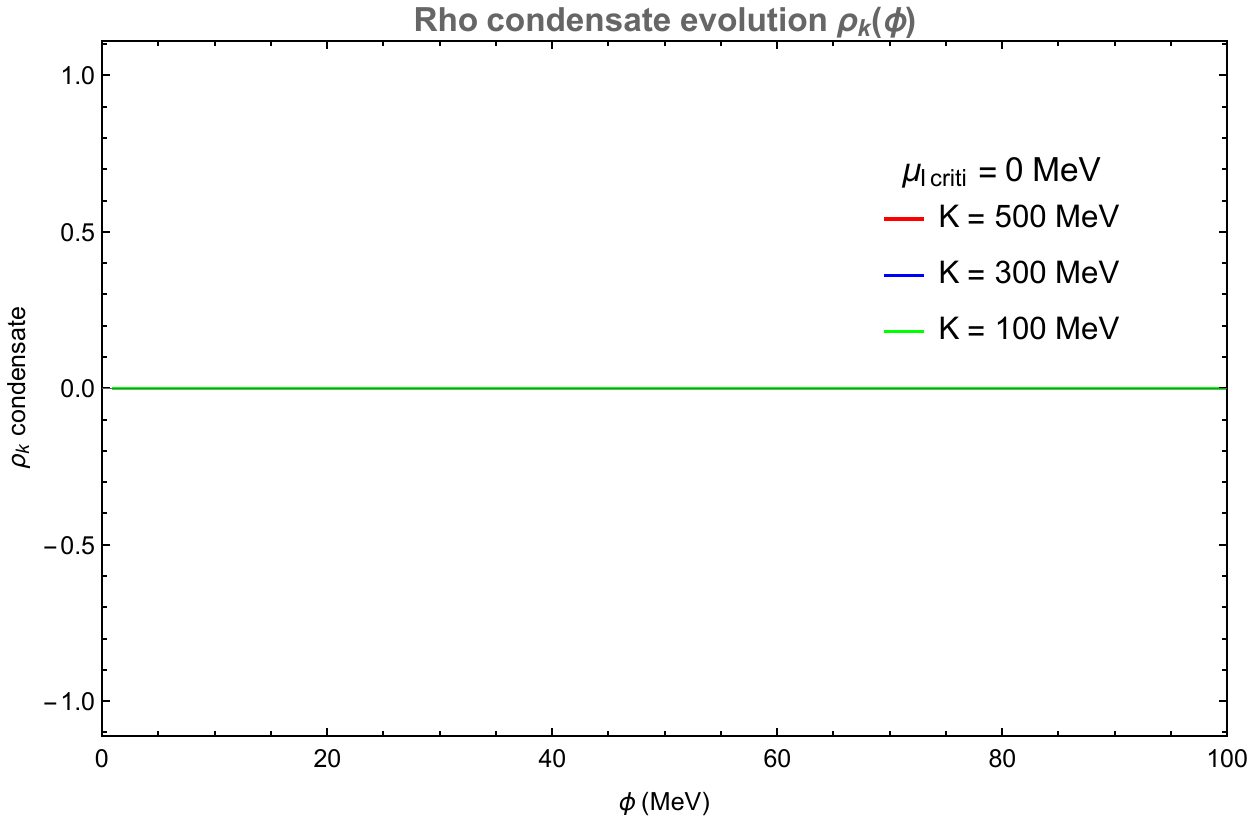} 
\caption{}
\label{}
\end{subfigure}
\caption{ (color online) $\rho$ condensate as a function of the field $\phi$ at  IR cutoff set to $20$ MeV and the ultraviolet cutoff set as $\Lambda_{FRG}$ =500 MeV: (a) $\mu = 100$ MeV, (b) $\mu = 200$ MeV, (c) $\mu = 240$ MeV, and (d) $\mu = 300$ MeV. The coupling was set as $\frac{{g_{\rho}}}{m_{\rho}} = 0.006$ MeV$^{-1}$ and the temperature was set as $T = 10$ MeV.}
\label{Expectation value for rho condensate}
\end{figure}
Fig.~\ref{Expectation value for rho condensate} shows the evolution of the $\rho$ condensate from microscopic to macroscopic scales under different isospin chemical potentials.
In the mean-field approximation, the condensation of the $\rho$ meson occurs only when the isospin chemical potential exceeds its vacuum mass, i.e., $\mu_I > m_\rho$. However, beyond the mean-field level, such as in the Random Phase Approximation (RPA) and Chiral Perturbation Theory (CPT) \cite{Brauner:2016lkh}, fluctuation effects significantly lower this critical chemical potential to $\mu_I = m_\pi$. This value agrees well with our own calculations in Figs ~\ref{The chiral and rho condensates as a function of isospin chemical (1)},~\ref{The chiral and rho condensates as a function of isospin chemical (2)} and ~\ref{The chiral and rho condensates as a function of isospin chemical (4)}. As shown in Fig.~\ref{Expectation value for rho condensate}, our results further confirm that when fluctuations are included, $\rho$ condensation appears at the IR scale. 
\begin{figure}[H]
\centering
\begin{subfigure}[b]{0.45\textwidth}
\includegraphics[width=\textwidth]{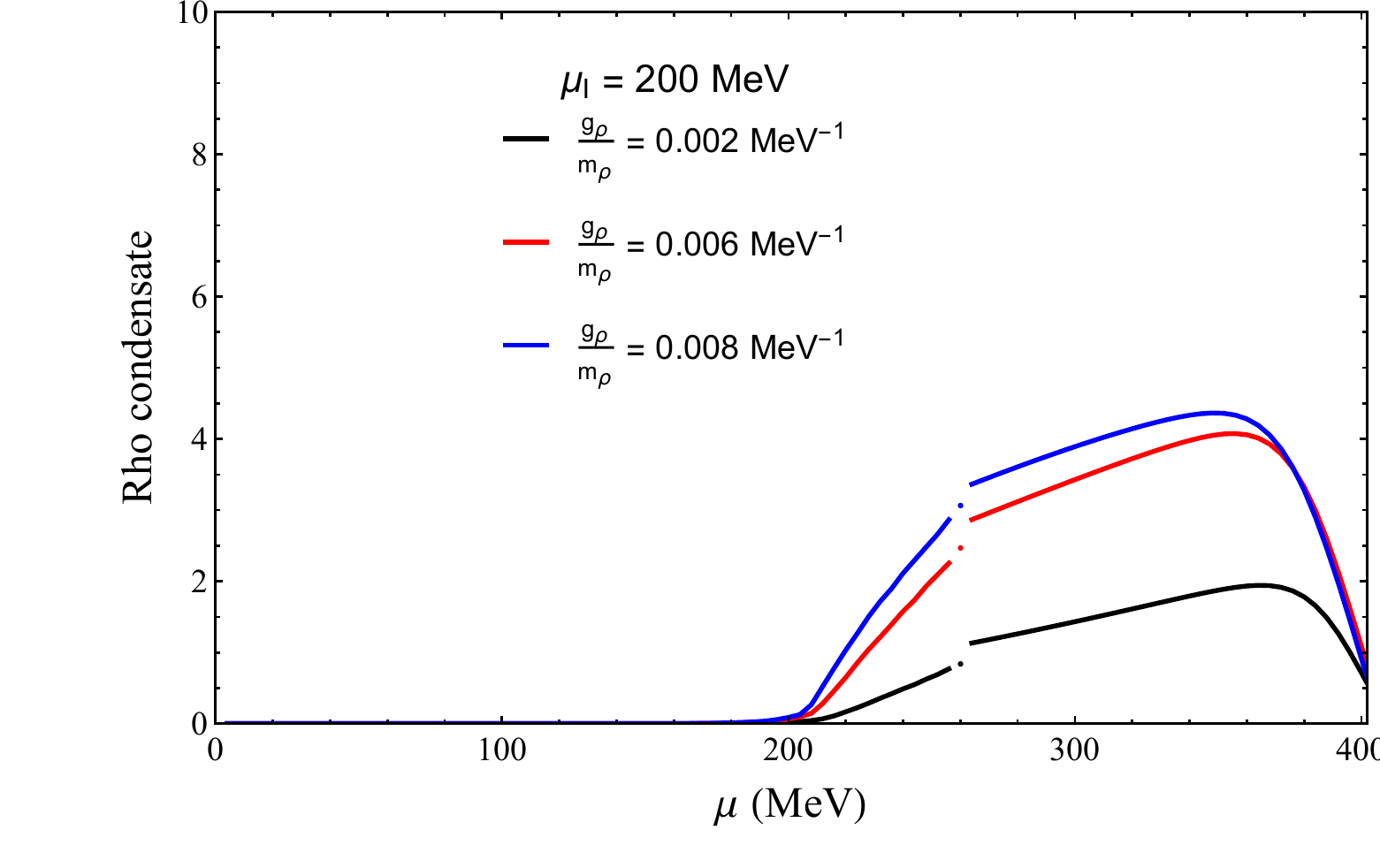} 
\caption{}
\end{subfigure}
\hfill
\begin{subfigure}[b]{0.45\textwidth}
\includegraphics[width=\textwidth]{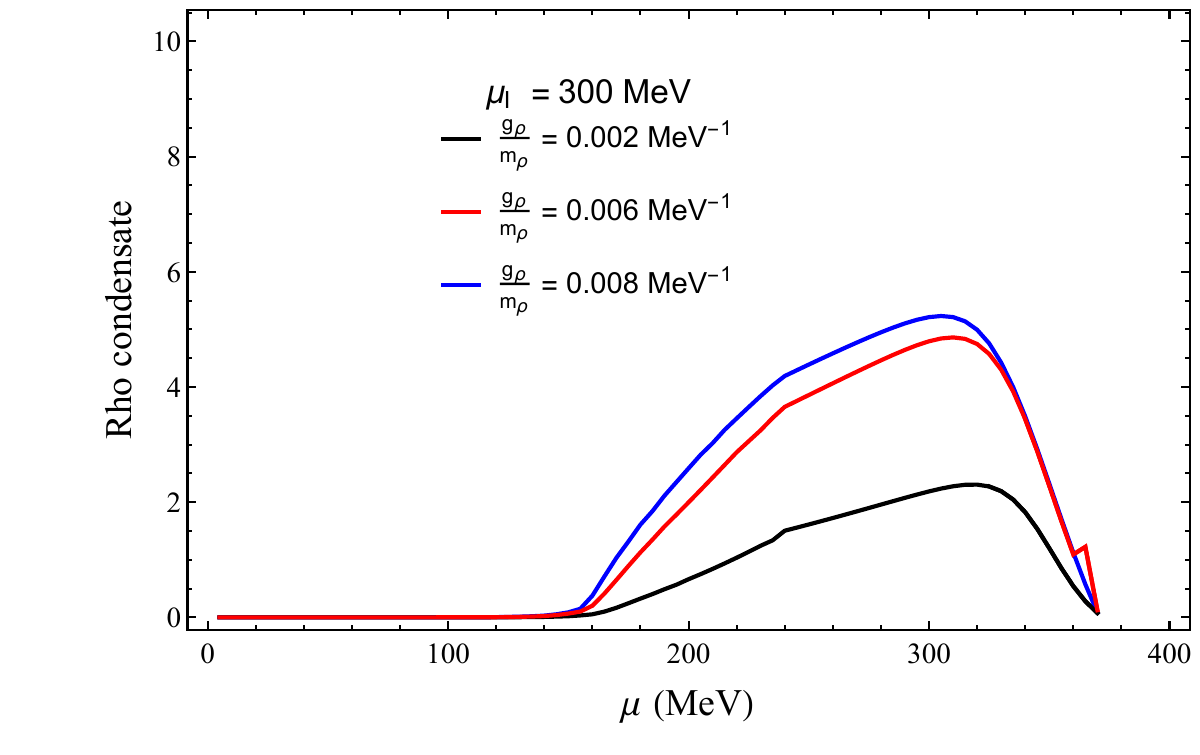} 
\caption{}
\end{subfigure}
\vskip\baselineskip
\begin{subfigure}[b]{0.45\textwidth}
\includegraphics[width=\textwidth]{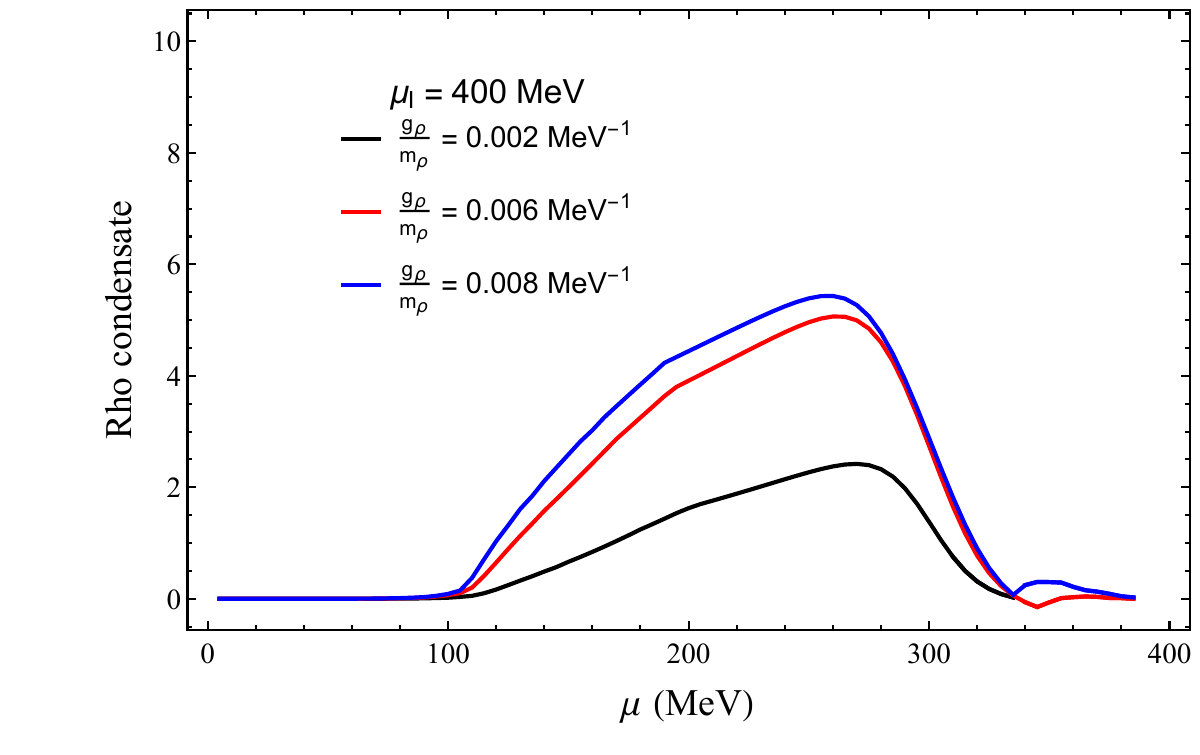}
\caption{}
\end{subfigure}
\hfill
\begin{subfigure}[b]{0.45\textwidth}
\includegraphics[width=\textwidth]{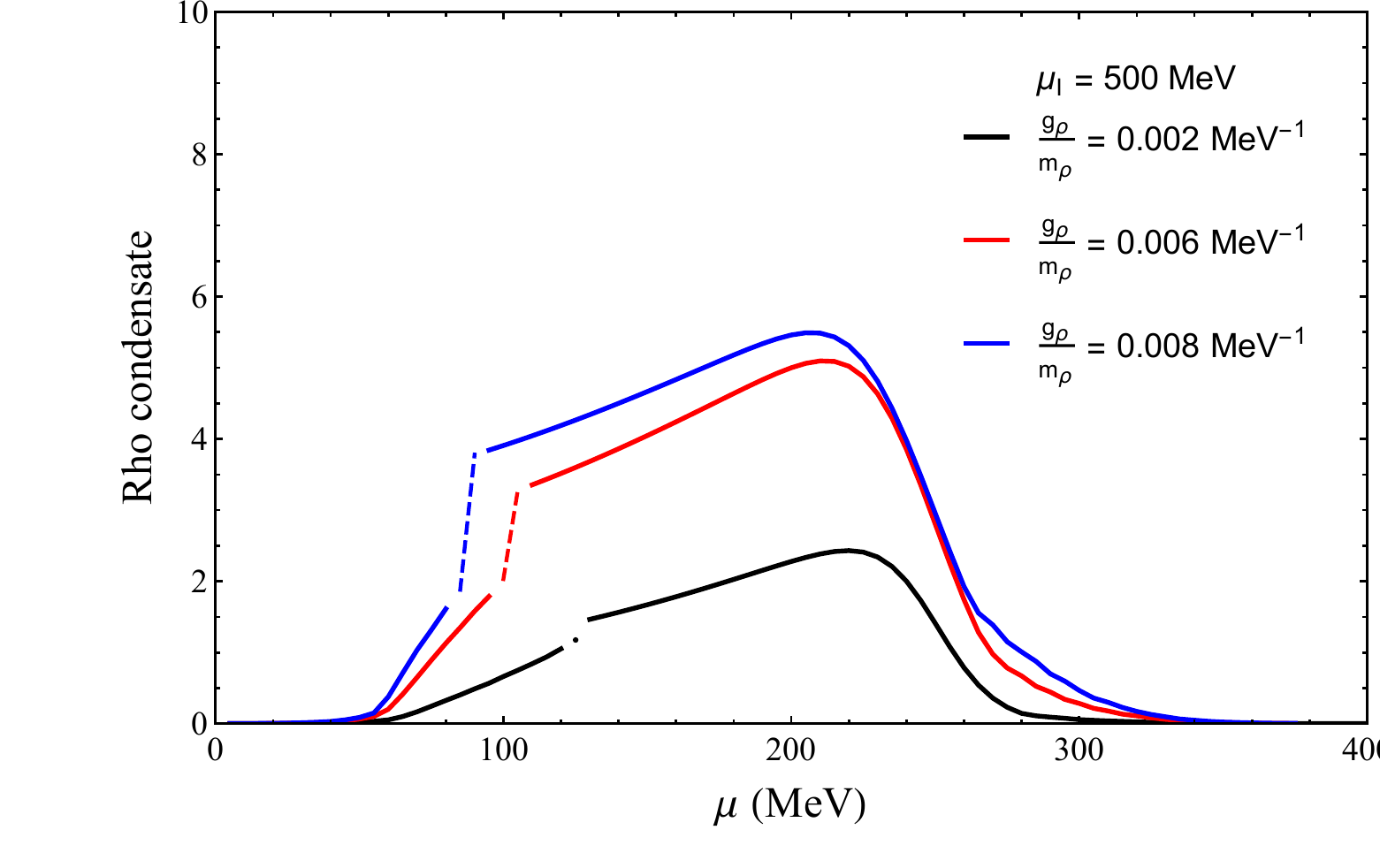} 
\caption{}
\end{subfigure}
\caption{ (color online) $\rho$ meson condensate as a function of the chemical potential for different values of the isospin chemical potential: (a) $\mu_{I} = 200$ MeV, (b) $\mu_{I} = 300$ MeV, (c) $\mu_{I} = 400$ MeV, and (d) $\mu_{I} = 500$ MeV. Here, $T = 10$ MeV.}
\label{fig: Rho con fn of chemicl potential}
\end{figure}
In Fig.~\ref{fig: Rho con fn of chemicl potential}, we show the $\rho$ condensate as a function of the chemical potential for different isospin chemical potentials: $\mu_{I} =  200$ MeV, $\mu_{I} =  300$ MeV,$\mu_{I}= 400$ MeV and $\mu_{I}= 500$ MeV, panels (a), (b), (c), and (d), respectively. Note that when the chemical potential increases, the $\rho$ condensate increases first and then drops to zero. Also the critical chemical potential shifts left when the isospin chemical potential increases at a fixed temperature. Furthermore, note that the $\rho$ condensate increases with increasing couplings at a fixed temperature.
Interestingly, the $\rho$ meson becomes condensate when the isospin chemical potential is larger then $210 \ \mathrm{MeV}$ for a fixed chemical potential $\mu = 200$ MeV; see panel (a) in Fig ~\ref{The chiral and rho condensates as a function of isospin chemical (2)}.
This is even more evident for a isospin chemical potential  $\mu_{I}$ = $400$ MeV (panel(c)) when the chemical potential is below $\mu$ = $100$ MeV, which is approximately the critical chemical potential for the $\rho$ meson condensate (in addition to the chiral condensate); see Fig~\ref{The chiral and rho condensates as a function of isospin chemical (1)}. This is case for all the isospin chemical potential relations with the chemical potential for the chiral and $\rho$ condensates.
\begin{figure}[H]
\centering
\begin{subfigure}[b]{0.45\textwidth}
\includegraphics[width=\textwidth]{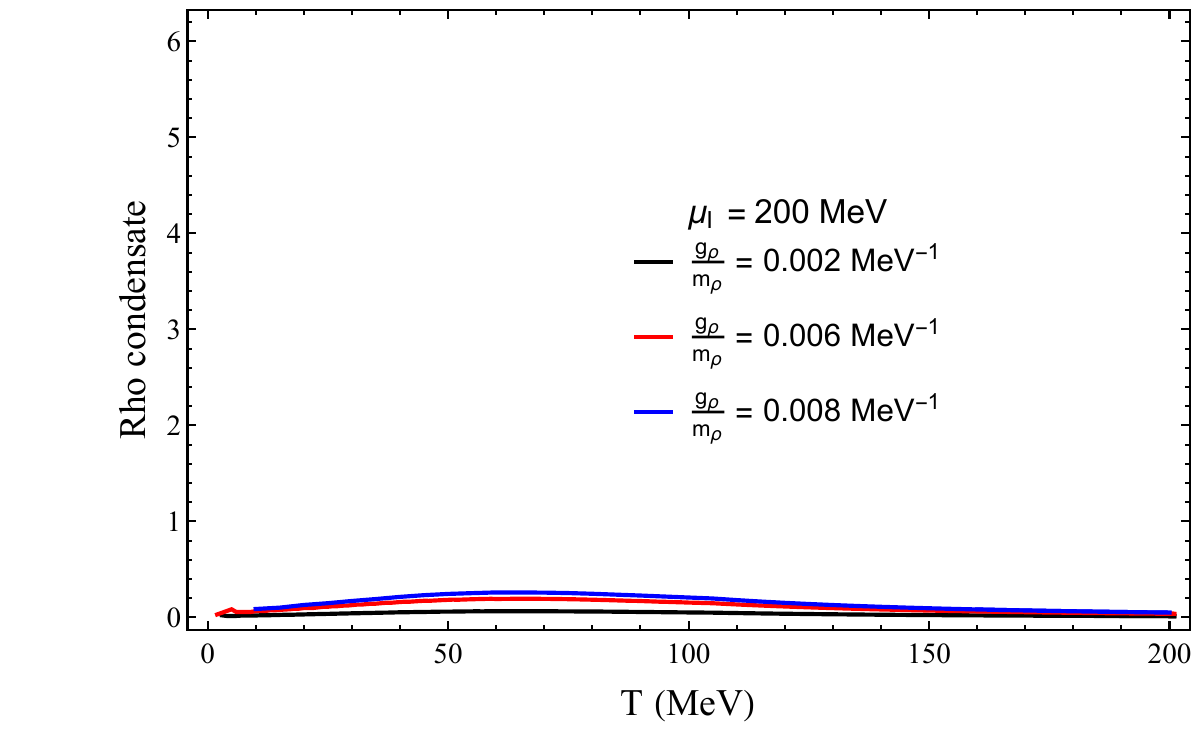} 
\caption{}
\end{subfigure}
\hfill
\begin{subfigure}[b]{0.45\textwidth}
\includegraphics[width=\textwidth]{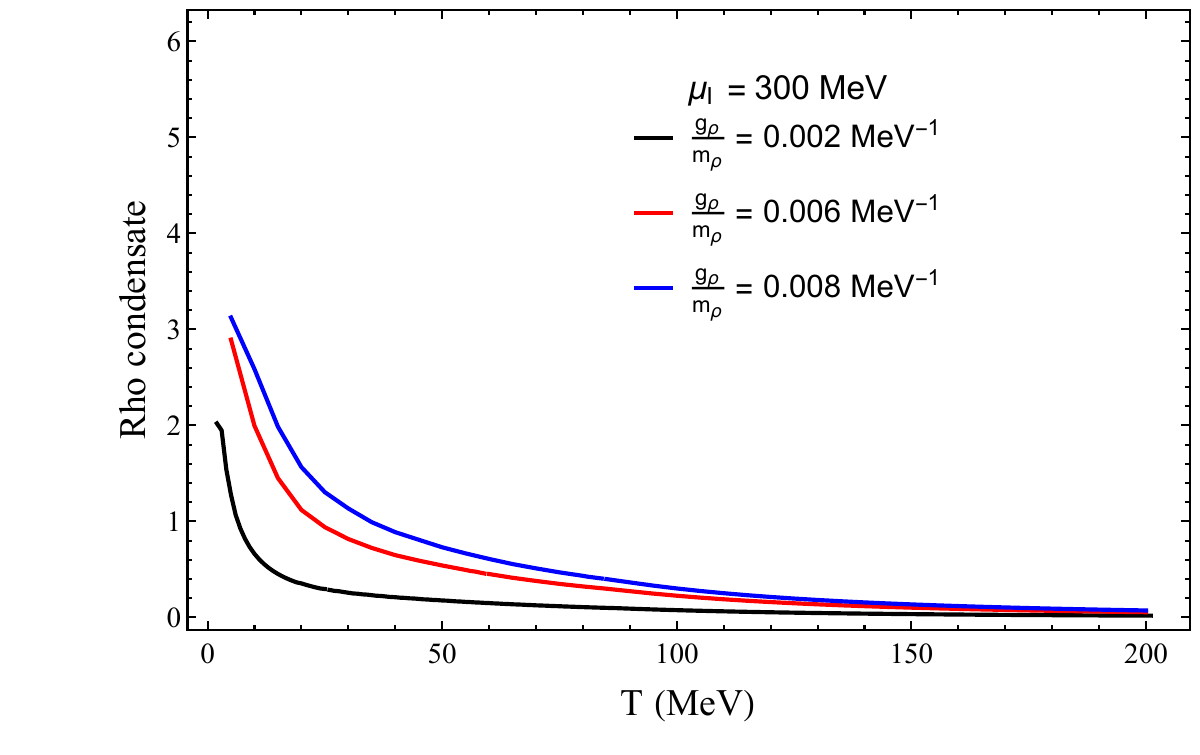} 
\caption{}
\end{subfigure}
\vskip\baselineskip
\begin{subfigure}[b]{0.45\textwidth}
\includegraphics[width=\textwidth]{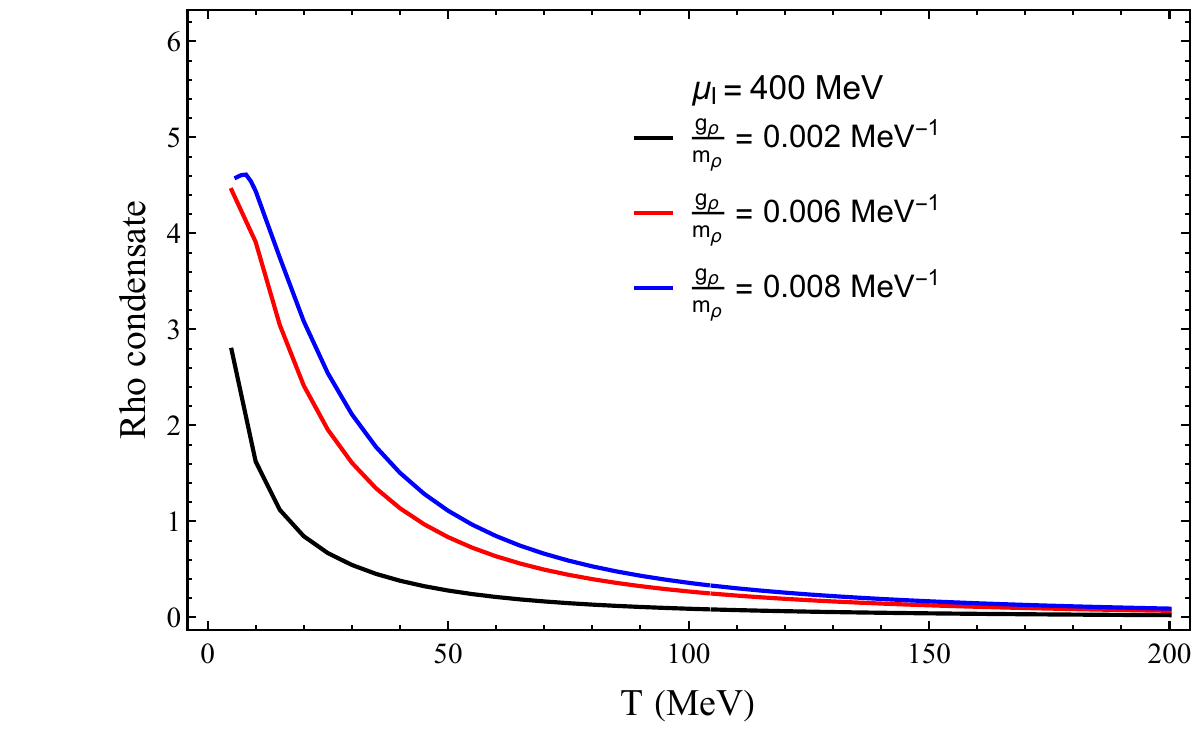}
\caption{}
\end{subfigure}
\hfill
\begin{subfigure}[b]{0.45\textwidth}
\includegraphics[width=\textwidth]{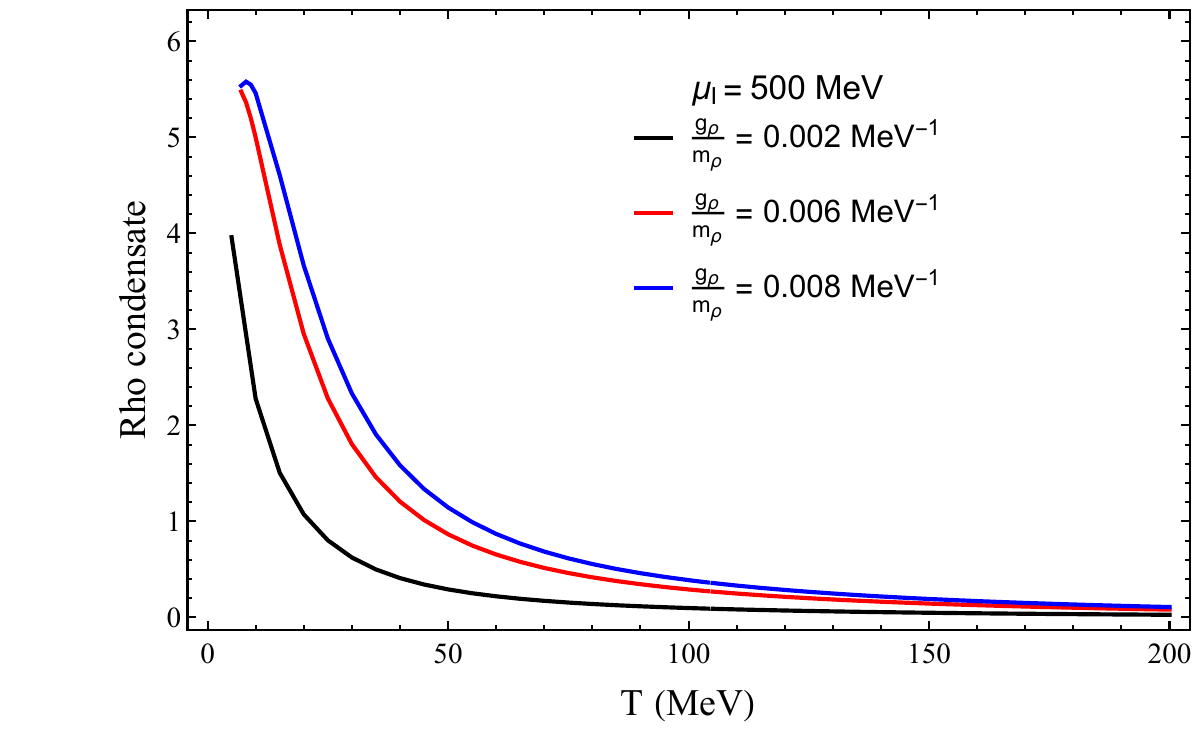} 
\caption{}
\end{subfigure}\\
\caption{ (color online) $\rho$ meson condensate as a function of temperature for different values of the isospin chemical potential: (a) $\mu_{I} = 200$ MeV, (b) $\mu_{I} = 300$ MeV, (c) $\mu_{I} = 400$ MeV, and (d) $\mu_{I} = 500$ MeV. Here, $\mu = 200$ MeV.}
\label{fig: Rho con fn of temperature}
\end{figure}
Finally, Fig.~\ref{fig: Rho con fn of temperature} shows the $\rho$ meson condensate as a function of temperature for different isospin chemical potentials: $\mu_{I} = 200~\mathrm{MeV}$ (panel (a)), $\mu_{I} = 400~\mathrm{MeV}$ (panel (b)), and $\mu_{I} = 500~\mathrm{MeV}$ (panel (c)), at a fixed $\mu = 200~\mathrm{MeV}$ and ${g}_{\rho}/m_{\rho} = 0.006~\mathrm{MeV}^{-1}$. The condensation value of the $\rho$ meson for $\mu_{I} = 200~\mathrm{MeV}$ (panel (a)) and $\mu = 200~\mathrm{MeV}$ is consistent with the results obtained in Fig.~\ref{The chiral and rho condensates as a function of isospin chemical (2)}, confirming the reliability of our calculations.
The coupling dependence shows that a larger $g_\rho$ enhances the magnitude of the $\rho$ condensate without altering its qualitative temperature dependence. At finite temperatures, the isovector order tends to melt, whereas stronger repulsion reinforces it at low $T$. This interplay explains why $\rho$ effects are most pronounced in cold, dense matter. As the temperature increases, the $\rho$ mode softens and approaches $\rho$–$a_1$ degeneracy, with the decreasing condensate following the chiral trend—signaling proximity to (partial) chiral restoration~\cite{Jung:2016yxl}.
\section{CONCLUSIONS}\label{sec:4}

We investigated the $\rho$ meson condensation in the isospin chemical potential by applying FRG, using a two-flavor quark-meson model including the $\rho$ meson. We also investigated the impact of vector mesons and isospin chemical potential on the phase structure of the chiral phase transition. The primary conclusions can be categorized into two facets: the impact of vector couplings and isospin chemical potential on the phase structure. The phase boundary moves as a unit to the low temperature and low-density region as the isospin chemical potential increases, with the vector coupling strength remaining constant. Consequently, the temperature of TCP gradually decreases. In contrast to changing only the vector coupling, as in ~\cite{CamaraPereira:2020xla}, the isospin chemical potential slightly reduces the temperature at which the phase transition occurs at low chemical potential, with no observed back bending behavior, consistent with the explanation provided in \cite{osman2025functional}.
In addition, we investigated the $\rho$ condensation as a function of the isospin chemical potential for different vector coupling constants. 
Beyond the mean-field approximation, fluctuation effects lower the critical isospin chemical potential for $\rho$ meson condensation from $\mu_I > m_\rho$ to $\mu_I = m_\pi$, at which the $\rho$ meson condenses alongside the chiral condensate, consistent with the results from RPA and CPT\cite{Brauner:2016lkh}, and confirmed by our FRG calculations.
We observed that at large chemical potential and large isospin chemical potential, the $\rho$ meson condensate dominates.
Increasing the coupling constant for the $\rho$ meson enhances the condensate value, though the critical isospin chemical potential remains relatively stable around 200 MeV. Increasing $\rho$ meson coupling slightly shifts the boundary of the phase transition.
\section{ACKNOWLEDGEMENT}

We thank Hai-cang Ren and Moran Jia for useful discussions. This work is supported in part by the National Key Research and
 Development Program of China under Contract
No. 2022YFA1604900. This work is also partly supported
by the National Natural Science Foundation of China
 (NSFC) under Grants No. 12435009, and No. 12275104.
 Hui Zhang acknowledges the financial support from the Guangdong
Major Project of Basic and Applied Basic Research (Grant No.
2020B0301030008) and the National Natural Science Foundation of China (Grant No. 12047523, and 12105107).
\section*{References}
\bibliographystyle{iopart-num}
\bibliography{References}
\end{document}